\PassOptionsToPackage{unicode}{hyperref}
\PassOptionsToPackage{hyphens}{url}
\PassOptionsToPackage{dvipsnames,svgnames,x11names}{xcolor}
\documentclass[
]{article}

\usepackage{amsmath,amssymb}
\usepackage{lmodern}
\usepackage{iftex}
\usepackage[margin=1.0in,bottom=1.5in]{geometry}
\ifPDFTeX
  \usepackage[T1]{fontenc}
  \usepackage[utf8]{inputenc}
  \usepackage{textcomp} 
\else 
  \usepackage{unicode-math}
  \defaultfontfeatures{Scale=MatchLowercase}
  \defaultfontfeatures[\rmfamily]{Ligatures=TeX,Scale=1}
\fi
\IfFileExists{upquote.sty}{\usepackage{upquote}}{}
\IfFileExists{microtype.sty}{
  \usepackage[]{microtype}
  \UseMicrotypeSet[protrusion]{basicmath} 
}{}
\makeatletter
\@ifundefined{KOMAClassName}{
  \IfFileExists{parskip.sty}{%
    \usepackage{parskip}
  }{
    \setlength{\parindent}{0pt}
    \setlength{\parskip}{6pt plus 2pt minus 1pt}}
}{
  \KOMAoptions{parskip=half}}
\makeatother
\usepackage{xcolor}
\ifx\paragraph\undefined\else
  \let\oldparagraph\paragraph
  \renewcommand{\paragraph}[1]{\oldparagraph{#1}\mbox{}}
\fi
\ifx\subparagraph\undefined\else
  \let\oldsubparagraph\subparagraph
  \renewcommand{\subparagraph}[1]{\oldsubparagraph{#1}\mbox{}}
\fi
\usepackage{lipsum}

\newcommand\blfootnote[1]{%
  \begingroup
  \renewcommand\thefootnote{}\footnote{#1}%
  \addtocounter{footnote}{-1}%
  \endgroup
}
\usepackage{color}
\usepackage{fancyvrb}

\DefineVerbatimEnvironment{Highlighting}{Verbatim}{commandchars=\\\{\}}
\usepackage{framed}
\definecolor{shadecolor}{RGB}{241,243,245}
\newenvironment{Shaded}{\begin{snugshade}}{\end{snugshade}}

\newcommand{\CharTok}[1]{\textcolor[rgb]{0.13,0.47,0.30}{#1}}
\newcommand{\CommentTok}[1]{\textcolor[rgb]{0.37,0.37,0.37}{#1}}

\newcommand{\ControlFlowTok}[1]{\textcolor[rgb]{0.00,0.23,0.31}{#1}}
\newcommand{\DataTypeTok}[1]{\textcolor[rgb]{0.68,0.00,0.00}{#1}}

\newcommand{\FloatTok}[1]{\textcolor[rgb]{0.68,0.00,0.00}{#1}}
\newcommand{\FunctionTok}[1]{\textcolor[rgb]{0.28,0.35,0.67}{#1}}

\newcommand{\KeywordTok}[1]{\textcolor[rgb]{0.00,0.23,0.31}{#1}}
\newcommand{\NormalTok}[1]{\textcolor[rgb]{0.00,0.23,0.31}{#1}}
\newcommand{\OperatorTok}[1]{\textcolor[rgb]{0.37,0.37,0.37}{#1}}

\providecommand{\tightlist}{%
  \setlength{\itemsep}{0pt}\setlength{\parskip}{0pt}}\usepackage{longtable,booktabs,array}
\usepackage{calc} 
\usepackage{etoolbox}
\makeatletter
\patchcmd\longtable{\par}{\if@noskipsec\mbox{}\fi\par}{}{}
\makeatother
\IfFileExists{footnotehyper.sty}{\usepackage{footnotehyper}}{\usepackage{footnote}}
\makesavenoteenv{longtable}
\usepackage{graphicx}
\makeatletter
\def\maxwidth{\ifdim\Gin@nat@width>\linewidth\linewidth\else\Gin@nat@width\fi}
\def\maxheight{\ifdim\Gin@nat@height>\textheight\textheight\else\Gin@nat@height\fi}
\makeatother
\setkeys{Gin}{width=\maxwidth,height=\maxheight,keepaspectratio}
\makeatletter
\def\fps@figure{htbp}
\makeatother
\newlength{\cslhangindent}
\newlength{\csllabelwidth}
\newlength{\cslentryspacingunit} 
\newenvironment{CSLReferences}[2] 
 {
  \setlength{\parindent}{0pt}
  \ifodd #1
  \let\oldpar\par
  \def\par{\hangindent=\cslhangindent\oldpar}
  \fi
  \setlength{\parskip}{#2\cslentryspacingunit}
 }%
 {}
\usepackage{calc}

\makeatletter
\@ifpackageloaded{caption}{}{\usepackage{caption}}
\AtBeginDocument{%
\ifdefined\contentsname
  \renewcommand*\contentsname{Table of contents}
\else
  \newcommand\contentsname{Table of contents}
\fi
\ifdefined\listfigurename
  \renewcommand*\listfigurename{List of Figures}
\else
  \newcommand\listfigurename{List of Figures}
\fi
\ifdefined\listtablename
  \renewcommand*\listtablename{List of Tables}
\else
  \newcommand\listtablename{List of Tables}
\fi
\ifdefined\figurename
  \renewcommand*\figurename{Figure}
\else
  \newcommand\figurename{Figure}
\fi
\ifdefined\tablename
  \renewcommand*\tablename{Table}
\else
  \newcommand\tablename{Table}
\fi
}
\@ifpackageloaded{float}{}{\usepackage{float}}
\floatstyle{ruled}
\@ifundefined{c@chapter}{\newfloat{codelisting}{h}{lop}}{\newfloat{codelisting}{h}{lop}[chapter]}
\floatname{codelisting}{Listing}

\makeatother
\makeatletter
\@ifpackageloaded{caption}{}{\usepackage{caption}}
\@ifpackageloaded{subcaption}{}{\usepackage{subcaption}}
\makeatother
\makeatletter
\@ifpackageloaded{tcolorbox}{}{\usepackage[many]{tcolorbox}}
\makeatother
\makeatletter
\@ifundefined{shadecolor}{\definecolor{shadecolor}{rgb}{.97, .97, .97}}
\makeatother
\ifLuaTeX
  \usepackage{selnolig}  
\fi
\IfFileExists{bookmark.sty}{\usepackage{bookmark}}{\usepackage{hyperref}}
\IfFileExists{xurl.sty}{\usepackage{xurl}}{} 
\hypersetup{
  pdftitle={Learned multiphysics inversion with differentiable programming and machine learning},
  pdfauthor={Mathias Louboutin*; Ziyi Yin*; Rafael Orozco; Thomas J. Grady II; Ali Siahkoohi; Gabrio Rizzuti; Philipp A. Witte; Olav Møyner; Gerard J. Gorman; Felix J. Herrmann; * First two authors contributed equally to this work.},
  colorlinks=true,
  linkcolor={blue},
  filecolor={Maroon},
  citecolor={Blue},
  urlcolor={Blue},
  pdfcreator={LaTeX via pandoc}}

\title{Learned multiphysics inversion with differentiable programming
and machine learning}
\author{Mathias Louboutin\textsuperscript{1,*}, Ziyi Yin\textsuperscript{1,*}, Rafael Orozco\textsuperscript{1}, Thomas J. Grady II\textsuperscript{1}, Ali Siahkoohi\textsuperscript{2},\\
\bigskip Gabrio Rizzuti\textsuperscript{3}, Philipp A. Witte\textsuperscript{4}, Olav Møyner\textsuperscript{5}, Gerard J. Gorman\textsuperscript{6}, Felix J. Herrmann\textsuperscript{1}\\
\textsuperscript{1} Georgia Institute of
Technology, \textsuperscript{2} Rice University,
\textsuperscript{3} University Medical Center Utrecht,\\\textsuperscript{4} Microsoft, \textsuperscript{5} SINTEF Digital, \textsuperscript{6} Imperial College London}
\date{}

\begin{document}
\maketitle
\blfootnote{\textsuperscript{*}These authors contributed equally. Corresponding author: mlouboutin3@gatech.edu}
\ifdefined\Shaded\renewenvironment{Shaded}{\begin{tcolorbox}[sharp corners, enhanced, boxrule=0pt, interior hidden, borderline west={3pt}{0pt}{shadecolor}, breakable, frame hidden]}{\end{tcolorbox}}\fi

\hypertarget{summary}{%
\subsection{Summary}\label{summary}}

We present the Seismic Laboratory for Imaging and Modeling/Monitoring
(\href{https://github.com/slimgroup}{SLIM}) open-source software
framework for computational geophysics and, more generally, inverse
problems involving the wave-equation (e.g., seismic and medical
ultrasound), regularization with learned priors, and learned neural
surrogates for multiphase flow simulations. By integrating multiple
layers of abstraction, our software is designed to be both readable and
scalable. This allows researchers to easily formulate their problems in
an abstract fashion while exploiting the latest developments in
high-performance computing. We illustrate and demonstrate our design
principles and their benefits by means of building a scalable prototype
for permeability inversion from time-lapse crosswell seismic data, which
aside from coupling of wave physics and multiphase flow, involves
machine learning.

\hypertarget{motivation}{%
\subsection{Motivation}\label{motivation}}

Thanks to major advancements in high-performance computing (HPC)
techniques, computational (exploration) geophysics has made giant leaps
over the past decades. These developments have, for instance, led to the
adoption of wave-equation-based inversion technologies such as
full-waveform inversion (FWI) and reverse-time migration (RTM) that,
thanks to their adherence to wave physics, have resulted in superior
imaging in complex geologies. While these techniques certainly rank
amongst the most sophisticated imaging technologies, their
implementation relies with few exceptions---most notably iWave++ (Sun
and Symes 2010), Julia Devito Inversion framework
(\href{https://github.com/slimgroup/JUDI.jl}{JUDI.jl} of the Seismic
Laboratory for Imaging and Modeling
(\href{https://slim.gatech.edu}{SLIM}), P. A. Witte, Louboutin, Kukreja,
et al. (2019); Mathias Louboutin et al. (2023)), and Chevron's
\href{https://github.com/ChevronETC/Examples}{COFII} (Washbourne et al.
2021)---on monolithic low-level (C/Fortran) implementations. As a
consequence, due to their lack of abstraction and modern programming
constructs, these low-level implementations are difficult and very
costly to maintain, especially when performance considerations prevail
over best software practices. While these implementation design choices
lead to performant code for specific problems, such as FWI, they often
hinder the implementation of new algorithms, e.g., based on different
objective functions or constraints, as well as coupling existing code
bases with external software libraries. For instance, combining
wave-equation-based inversion with machine learning frameworks or
coupling wave-physics with multiphase fluid-flow solvers are considered
challenging and costly. Thus, our industry runs the risk of losing its
ability to innovate, a situation that is exacerbated by the challenges
we face as a result of the energy transition.

\hypertarget{design-principles}{%
\subsection{Design principles}\label{design-principles}}

To address these important shortcomings of current software
implementations that impede progress, we have embarked on the
development of a performant software framework. For instance, our wave
propagators, implemented in
\href{https://github.com/devitocodes/devito}{Devito} (M. Louboutin et
al. 2019; Luporini et al. 2020), are used in production by contractors
and Oil \& Gas majors while enabling rapid, low-cost, scalable, and
interoperable algorithm development for multiphysics and machine
learning problems that runs on a variety of different chipsets (e.g.,
ARM, Intel, POWER) and graphics accelerators (e.g., NVIDIA). To achieve
this, we adopt contemporary software design practices that include
high-level abstractions, software design principles, and the utilization
of modern programming languages, such as Python (Rossum and Drake 2009)
and \href{https://julialang.org}{Julia} (Bezanson et al. 2017). We also
make extensive use of abstractions provided by domain-specific languages
(DSLs), such as the Rice Vector Library (RVL, Padula, Scott, and Symes
2009) and the Unified Form Language (UFL, Rathgeber et al. 2016; Alnaes
et al. 2015), and adopt reproducible research practices introduced by
the trailblazing open-source initiative Madagascar (Fomel et al. 2013),
which successfully made use of version control and an abstraction based
on the software construction tool \texttt{SCons}.

In an effort to meet the challenges of modern software design in a
performance-critical environment, we adhere to three key principles---in
addition to the fundamental principle of separation of concerns. First,
we adopt mathematical language to inform our abstractions. Mathematics
is concise, unambiguous, well understood, and leads to natural
abstractions for the

\begin{itemize}
\tightlist
\item
  \textbf{wave physics}, through partial differential equations as put
  to practice by \href{https://github.com/devitocodes/devito}{Devito},
  which relies on Symbolic Python
  (\href{https://www.sympy.org/en/index.html}{SymPy}) (Meurer et al.
  2017) to define partial differential equations. Given the symbolic
  expressions, \href{https://github.com/devitocodes/devito}{Devito}
  automatically generates highly-optimized, possibly domain-decomposed,
  parallel C code that targets the available hardware with near-optimal
  performance for 3D acoustic, tilted-transverse-isotropic, or elastic
  wave-equations;
\item
  \textbf{linear algebra}, through matrix-free linear operators, as in
  \href{https://github.com/slimgroup/JUDI.jl}{JUDI.jl} (P. A. Witte,
  Louboutin, Kukreja, et al. 2019; Mathias Louboutin et al. 2023)---a
  high-level linear algebra DSL for wave-equation-based modeling and
  inversion. These ideas date back to
  \href{https://github.com/mpf/spot}{SPOT} (Berg and Friedlander 2009)
  with more recent implementations
  \href{https://github.com/slimgroup/JOLI.jl}{JOLI.jl} (Modzelewski et
  al. 2023) in Julia and
  \href{https://pylops.readthedocs.io/en/stable/}{PyLops} in Python
  (Ravasi and Vasconcelos 2020).
\item
  \textbf{optimization}, through definition of objective functions, also
  known as loss functions, that need to be minimized---via
  \href{https://github.com/slimgroup/SlimOptim.jl}{SlimOptim.jl}
  (Mathias Louboutin, Yin, and Herrmann 2022b)---subject to mathematical
  constraints, which can be imposed through
  \href{https://github.com/slimgroup/SetIntersectionProjection.jl}{SetIntersectionProjection.jl}
  (Peters and Herrmann 2019; Peters, Louboutin, and Modzelewski 2022).
\end{itemize}

Second, we exploit hierarchy within wave-equation-based inversion
problems that naturally leads to a separation of concerns. At the
highest level, we deal with linear operators, specifically matrix-free
Jacobians of wave-based inversion, with
\href{https://github.com/slimgroup/JUDI.jl}{JUDI.jl} and parallel file
input/output with
\href{https://github.com/slimgroup/SegyIO.jl}{SegyIO.jl} (Lensink et al.
2023) on premise, or in the Cloud (Azure) via
\href{https://github.com/slimgroup/JUDI4Cloud.jl}{JUDI4Cloud.jl}
(Mathias Louboutin, Yin, and Herrmann 2022a). At the intermediate and
lower level, we make extensive use of
\href{https://github.com/devitocodes/devito}{Devito} (M. Louboutin et
al. 2019; Luporini et al. 2020)---a just-in-time compiler for
stencil-based time-domain finite-difference calculations, the
development of which \href{https://slim.gatech.edu}{SLIM} has been
involved in over the years.

Third, we build on the principles of differentiable programming as
advocated by Mike Innes et al. (2019) and intrusive automatic
differentiation introduced by D. Li et al. (2020) to integrate
wave-physics with machine learning frameworks and multiphase flow.
Specifically, we employ automatic differentiation (AD) through the use
of the chain rule, including abstractions that allow the user to add
derivative rules, as in
\href{https://github.com/JuliaDiff/ChainRules.jl}{ChainRules.jl} (White
et al. 2022, 2023).

During the Inaugural Full-Waveform Inversion Workshop in 2015, we at
\href{https://slim.gatech.edu}{SLIM} started to articulate these design
principles (Lin and Herrmann 2015), which over the years cumulated in
scalable parallel software frameworks for time-harmonic FWI (Silva and
Herrmann 2019), for time-domain RTM and FWI (P. A. Witte, Louboutin,
Kukreja, et al. 2019; Mathias Louboutin et al. 2023), and for abstracted
FWI (Mathias Louboutin et al. 2022) allowing for connections with
machine learning. Aside from developing software for wave-equation-based
inversion, our group has more recently also been involved in the
development of scalable machine learning solutions, including the Julia
package
\href{https://github.com/slimgroup/InvertibleNetworks.jl}{InvertibleNetworks.jl}
(P. Witte et al. 2023, 2023), which implements memory-efficient
invertible deep neural networks such as (conditional) normalizing flows
(NFs, Rezende and Mohamed 2015), and scalable distributed Fourier neural
operators (FNOs, Z. Li et al. 2020) in the
\href{https://github.com/slimgroup/dfno}{dfno} package (Grady et al.
2022; Grady, Infinoid, and Louboutin 2022). All of these will be
described in more detail below.

\begin{figure}

{\centering 

\includegraphics{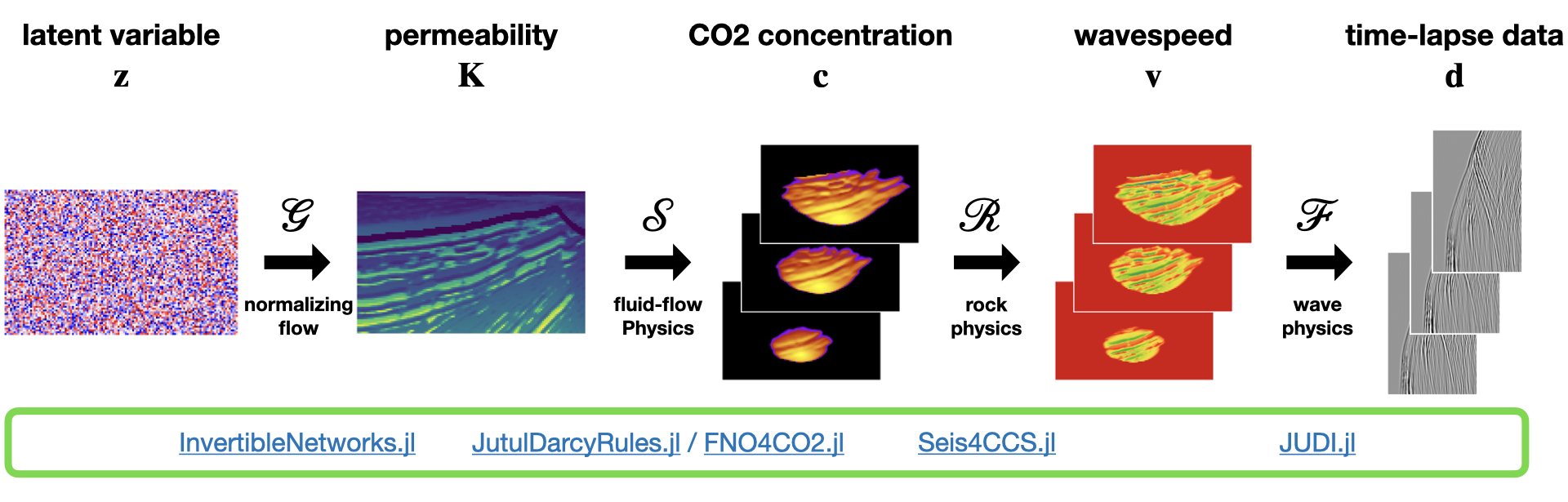}

}

\caption{\label{fig-end2end-latent}The multiphysics forward model. The
permeability, \(\mathbf{K}\), is generated from Gaussian noise with a
pretrained NF, \(\mathcal{G}\), followed by two-phase flow simulations
through \(\mathcal{S}\), rock physics denoted by \(\mathcal{R}\), and
time-lapse seismic data simulations via wave physics, \(\mathcal{F}\).}

\end{figure}

To illustrate how these design principles can lead to solutions of
complex learned coupled inversions, we consider in the ensuing sections
end-to-end inversion of time-lapse seismic data for the spatial
permeability distribution (D. Li et al. 2020). As can be seen from
Figure~\ref{fig-end2end-latent}, this inversion problem is rather
complex and whose solution arguably benefits from our three design
principles listed above. In this formulation, the latent representation
for the permeability is taken via a series of nonlinear operations all
the way to the time-lapse seismic data. In the remainder of this
exposition, we will detail how the different components in this learned
inversion problem are implemented so that the coupled inversion can be
carried out. The results presented are preliminary representing a
snapshot on how research is conducted according to the design
principles.

\hypertarget{learned-time-lapse-end-to-end-permeability-inversion}{%
\subsection{Learned time-lapse end-to-end permeability
inversion}\label{learned-time-lapse-end-to-end-permeability-inversion}}

\begin{figure}

{\centering 

\includegraphics[width=0.5\textwidth,height=\textheight]{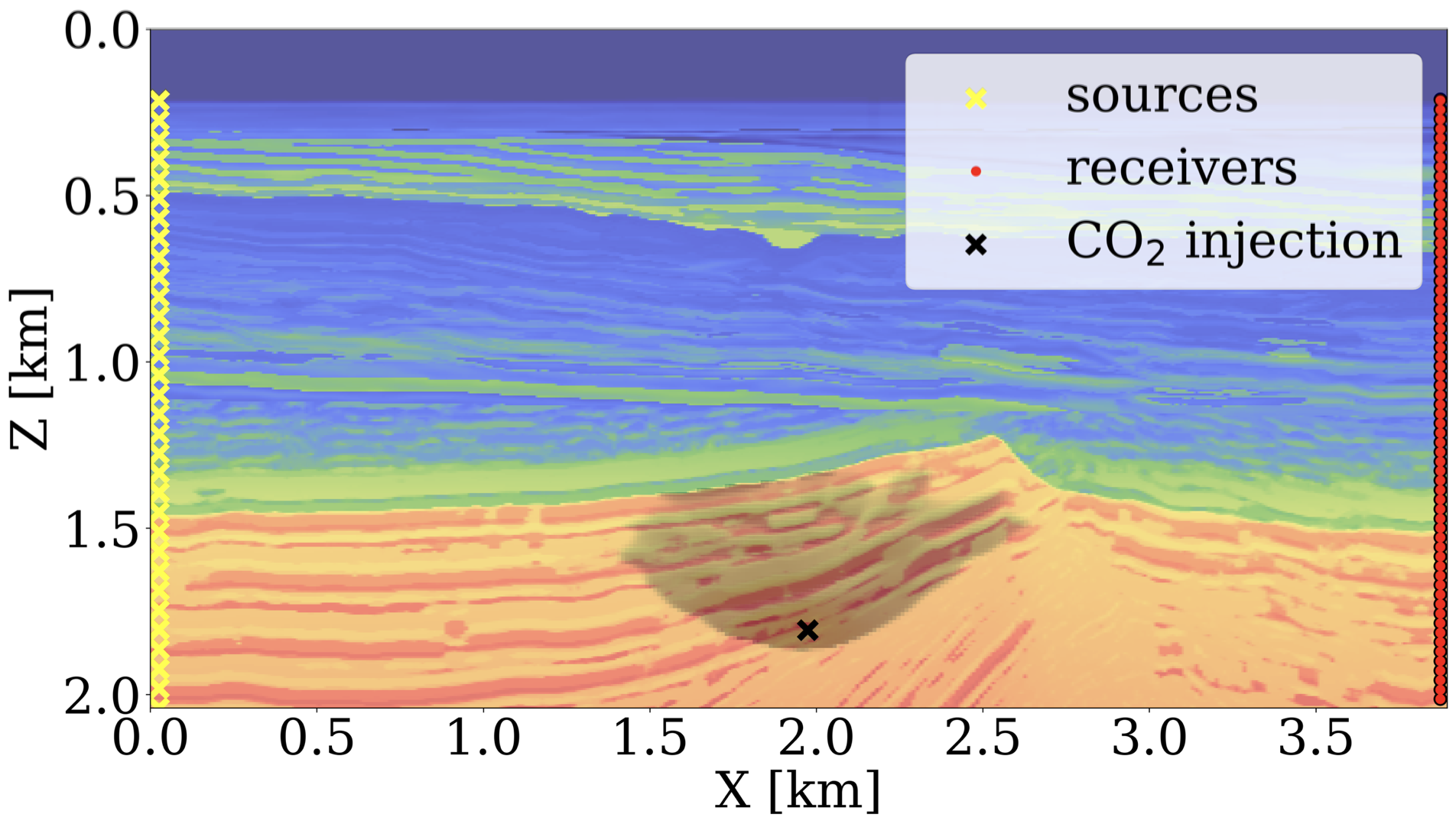}

}

\caption{\label{fig-end2end-acquisition}Experimental setup. The black
\(\times\) symbol in the middle of the model indicates the
CO\textsubscript{2} injection location. The seismic sources are on the
left-hand side of the model (shown as yellow \(\times\) symbols) and
receivers are on the right-hand side of the model (shown as red dots).
In grey color, we overlay the compressional wavespeed with simulated
CO\textsubscript{2} saturation modeled for 18 years.}

\end{figure}

Combating climate change and dealing with the energy transition call for
solutions to problems of increasing complexity. Building seismic
monitoring systems for geological CO\textsubscript{2} and/or
H\textsubscript{2} storage falls in this category. To demonstrate how
math-inspired abstractions can help, we consider inversion of
permeability from crosswell time-lapse data (see
Figure~\ref{fig-end2end-acquisition} for experimental setup) involving
(i) coupling of wave physics with two-phase (brine/CO\textsubscript{2})
flow using \href{https://github.com/sintefmath/Jutul.jl}{Jutul.jl}
(Møyner et al. 2023), state-of-the-art reservoir modeling software in
Julia; (ii) learned regularization with NFs with
\href{https://github.com/slimgroup/InvertibleNetworks.jl}{InvertibleNetworks.jl};
(iii) learned surrogates for the fluid-flow simulations with FNOs. This
type of inversion problem is especially challenging because it involves
different types of physics to estimate the past, current, and future
saturation and pressure distributions of CO\textsubscript{2} plumes from
crosswell data in saline aquifers. In the subsequent sections, we
demonstrate how we invert time-lapse data using the separate software
packages listed in Figure~\ref{fig-end2end-latent}.

\hypertarget{wave-equation-based-inversion}{%
\subsubsection{Wave-equation-based
inversion}\label{wave-equation-based-inversion}}

Thanks to its unmatched ability to resolve CO\textsubscript{2} plumes,
active-source time-lapse seismic is arguably the preferred imaging
modality when monitoring geological storage (Ringrose 2020). In its
simplest form for a single time-lapse vintage, FWI involves minimizing
the \(\ell_2\)-norm misfit/loss function between observed and synthetic
data---i.e., we have

\begin{equation}\protect\hypertarget{eq-fwi}{}{
\underset{\mathbf{m}}{\operatorname{minimize}} \quad \frac{1}{2}\|\mathbf{F}(\mathbf{m})\mathbf{q} - \mathbf{d} \|^2_2 \quad \text{where} \quad \mathbf{F}(\mathbf{m}) = \mathbf{P}_r {\mathbf{A}(\mathbf{m})}^{-1}\mathbf{P}_s^{\top}.
}\label{eq-fwi}\end{equation}

In this formulation, the symbol \(\mathbf{F}(\mathbf{m})\) represents
the forward modeling operator (wave physics), parameterized by the
squared slowness, \(\mathbf{m}\). This forward operator acting on the
sources consists of the composition of source injection operator,
\(\mathbf{P}^\top_s\) with \(^\top\) denoting the transpose operator,
solution of the discretized wave equation via
\({\mathbf{A}(\mathbf{m})}^{-1}\), and restriction to the receivers via
the linear operator \(\mathbf{P}_r\). The vector \(\mathbf{q}\)
represents the seismic sources and the vector \(\mathbf{d}\) contains
single-vintage seismic data, collected at the receiver locations. Thanks
to our adherence to the math, the corresponding Julia code to invert for
the unknown squared slowness \(\mathbf{m}\) with
\href{https://github.com/slimgroup/JUDI.jl}{JUDI.jl} reads

\begin{Shaded}
\begin{Highlighting}[]
\CommentTok{\# Forward modeling to generate seismic data.}
\NormalTok{Pr }\OperatorTok{=} \FunctionTok{judiProjection}\NormalTok{(recGeometry)  }\CommentTok{\# setup receiver}
\NormalTok{Ps }\OperatorTok{=} \FunctionTok{judiProjection}\NormalTok{(srcGeometry)  }\CommentTok{\# setup sources}
\NormalTok{Ainv }\OperatorTok{=} \FunctionTok{judiModeling}\NormalTok{(model)        }\CommentTok{\# setup wave{-}equation solver}
\NormalTok{F }\OperatorTok{=}\NormalTok{ Pr }\OperatorTok{*}\NormalTok{ Ainv }\OperatorTok{*}\NormalTok{ Ps}\OperatorTok{\textquotesingle{}}               \CommentTok{\# forward modeling operator}
\NormalTok{d }\OperatorTok{=} \FunctionTok{F}\NormalTok{(m\_true) }\OperatorTok{*}\NormalTok{ q                 }\CommentTok{\# generate observed data}
\CommentTok{\# Gradient descent to invert for the unknown squared slowness.}
\ControlFlowTok{for}\NormalTok{ it }\OperatorTok{=} \FloatTok{1}\OperatorTok{:}\NormalTok{maxiter}
\NormalTok{    d0 }\OperatorTok{=} \FunctionTok{F}\NormalTok{(m) }\OperatorTok{*}\NormalTok{ q                 }\CommentTok{\# generate synthetic data}
\NormalTok{    J }\OperatorTok{=} \FunctionTok{judiJacobian}\NormalTok{(}\FunctionTok{F}\NormalTok{(m), q)     }\CommentTok{\# setup the Jacobian operator of F}
\NormalTok{    g }\OperatorTok{=}\NormalTok{ J}\OperatorTok{\textquotesingle{}} \OperatorTok{*}\NormalTok{ (d0 }\OperatorTok{{-}}\NormalTok{ d)             }\CommentTok{\# gradient w.r.t. squared slowness}
\NormalTok{    m }\OperatorTok{=}\NormalTok{ m }\OperatorTok{{-}}\NormalTok{ t }\OperatorTok{*}\NormalTok{ g                 }\CommentTok{\# gradient descent with steplength t}
\ControlFlowTok{end}
\end{Highlighting}
\end{Shaded}

To obtain this concise and abstract formulation for FWI, we utilized
hierarchical abstractions for propagators in
\href{https://github.com/devitocodes/devito}{Devito} and linear algebra
tools in \href{https://github.com/slimgroup/JUDI.jl}{JUDI.jl}, including
matrix-free implementations for \texttt{F} and its Jacobian \texttt{J}.
While the above stand-alone implementation allows for
(sparsity-promoting) seismic (P. A. Witte, Louboutin, Luporini, et al.
2019; Herrmann, Siahkoohi, and Rizzuti 2019; Yang et al. 2020; Rizzuti
et al. 2020, 2021; Siahkoohi, Rizzuti, and Herrmann 2020a, 2020b, 2020c;
Yin, Louboutin, and Herrmann 2021; Yin et al. 2023) and medical (Yin et
al. 2020; Orozco et al. 2021, 2023; Orozco, Louboutin, et al. 2023)
inversions, it relies on hand-derived implementations for the adjoint of
the Jacobian \texttt{J\textquotesingle{}} and for the derivative of the
loss function. Although this approach is viable, relying solely on
hand-derived derivatives can become cumbersome when we want to utilize
machine learning models or when we need to couple the wave equation to
the multiphase flow equation.

To allow for this situation, we make use of Julia's differentiable
programming ecosystem that includes tools to use AD and to add
differentiation rules via
\href{https://github.com/JuliaDiff/ChainRules.jl}{ChainRules.jl}. Using
this tool, the AD system can be taught how to differentiate
\href{https://github.com/slimgroup/JUDI.jl}{JUDI.jl} via the following
differentiation rule for the forward propagator

\begin{Shaded}
\begin{Highlighting}[]
\CommentTok{\# Custom AD rule for wave modeling operator.}
\KeywordTok{function} \FunctionTok{rrule}\NormalTok{(}\OperatorTok{::}\DataTypeTok{typeof}\NormalTok{(}\OperatorTok{*}\NormalTok{), F}\OperatorTok{::}\DataTypeTok{judiModeling}\NormalTok{, q)}
\NormalTok{    y }\OperatorTok{=}\NormalTok{ F }\OperatorTok{*}\NormalTok{ q                     }\CommentTok{\# forward modeling}
    \CommentTok{\# The pullback function for gradient calculations.}
    \FunctionTok{pullback}\NormalTok{(dy) }\OperatorTok{=} \FunctionTok{NoTangent}\NormalTok{(), }\FunctionTok{judiJacobian}\NormalTok{(F, q)}\CharTok{\textquotesingle{} * dy, F\textquotesingle{}} \OperatorTok{*}\NormalTok{ dy }
    \ControlFlowTok{return}\NormalTok{ y, pullback}
\KeywordTok{end}
\end{Highlighting}
\end{Shaded}

In this rule, the \texttt{pullback} function takes as input the data
residual, \texttt{dy}, and outputs the gradient with respect to the
operator \texttt{*} (no gradient), the model parameters, and the source
distribution. With this differentiation rule, the above gradient descent
algorithm can be implemented as follows:

\begin{Shaded}
\begin{Highlighting}[]
\CommentTok{\# Define the loss function.}
\FunctionTok{loss}\NormalTok{(m) }\OperatorTok{=} \FloatTok{.5f0} \OperatorTok{*} \FunctionTok{norm}\NormalTok{(}\FunctionTok{F}\NormalTok{(m) }\OperatorTok{*}\NormalTok{ q }\OperatorTok{{-}}\NormalTok{ d)}\OperatorTok{\^{}}\FloatTok{2f0}
\CommentTok{\# Gradient descent to invert for the squared slowness.}
\ControlFlowTok{for}\NormalTok{ it }\OperatorTok{=} \FloatTok{1}\OperatorTok{:}\NormalTok{maxiter}
\NormalTok{    g }\OperatorTok{=} \FunctionTok{gradient}\NormalTok{(loss, m)[}\FloatTok{1}\NormalTok{]      }\CommentTok{\# gradient computation via AD}
\NormalTok{    m }\OperatorTok{=}\NormalTok{ m }\OperatorTok{{-}}\NormalTok{ t }\OperatorTok{*}\NormalTok{ g                 }\CommentTok{\# gradient descent with steplength t}
\ControlFlowTok{end}
\end{Highlighting}
\end{Shaded}

Compared to the original implementation, this code only needs
\texttt{F(m)} and the function \texttt{loss(m)}. With the help of the
above \texttt{rrule}, Julia's AD system\footnote{In this case, we used
  reverse AD provided by
  \href{https://github.com/FluxML/Zygote.jl}{Zygote.jl}, the AD system
  provided by \href{https://julialang.org}{Julia} machine learning
  package \href{https://github.com/FluxML/Flux.jl}{Flux.jl}. Because
  \href{https://github.com/JuliaDiff/ChainRules.jl}{ChainRules.jl} is AD
  system agnostic, another choice could have been made.} is capable of
computing the gradients (line 5). Aside from remaining performant, i.e.,
we still make use of the adjoint-state method to compute the gradients,
the advantage of this approach is that it allows for much more
flexibility, e.g., in situations where the squared slowness is
parameterized in terms of a pretrained neural network or in terms of the
output of multiphase flow simulations. In the next section, we show how
trained NFs can serve as priors to improve the quality of FWI.

\hypertarget{deep-priors-and-normalizing-flows}{%
\subsubsection{Deep priors and normalizing
flows}\label{deep-priors-and-normalizing-flows}}

NFs are generative models that take advantage of invertible deep neural
network architectures to learn complex distributions from training
examples (Dinh, Sohl-Dickstein, and Bengio 2016). For example, in
seismic inversion applications, we are interested in approximating the
distribution of Earth models to use as priors in downstream tasks. NFs
learn to map samples from the target distribution (i.e., Earth models)
to zero-mean unit standard deviation Gaussian noise using a sequence of
trainable nonlinear invertible layers. Once trained, one can resample
new Gaussian noise and pass it through the inverse sequence of layers to
obtain new generative samples from the target distribution. NFs are an
attractive choice for generative models in seismic applications (Zhang
and Curtis 2020, 2021; Zhao, Curtis, and Zhang 2021; Siahkoohi and
Herrmann 2021; Siahkoohi et al. 2021, 2023; Siahkoohi, Rizzuti, and
Herrmann 2022) because they provide fast sampling and allow for
memory-efficient training due to their intrinsic invertibility, which
eliminates the need to store intermediate activations during
backpropagation. Memory efficiency is particularly important for seismic
applications due to the 3D volumetric nature of the seismic models.
Thus, our methods need to scale well in this regime.

\begin{figure}

{\centering 

\includegraphics{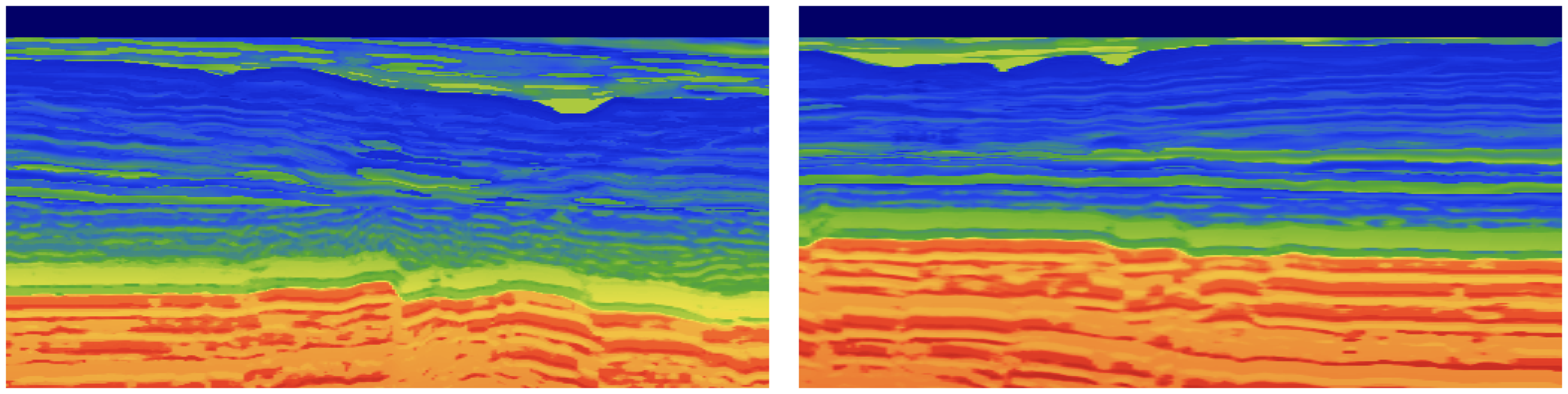}

}

\caption{\label{fig-compasssamples}Examples of Compass 2D slices used to
train a normalizing flow prior.}

\end{figure}

\begin{figure}

{\centering 

\includegraphics{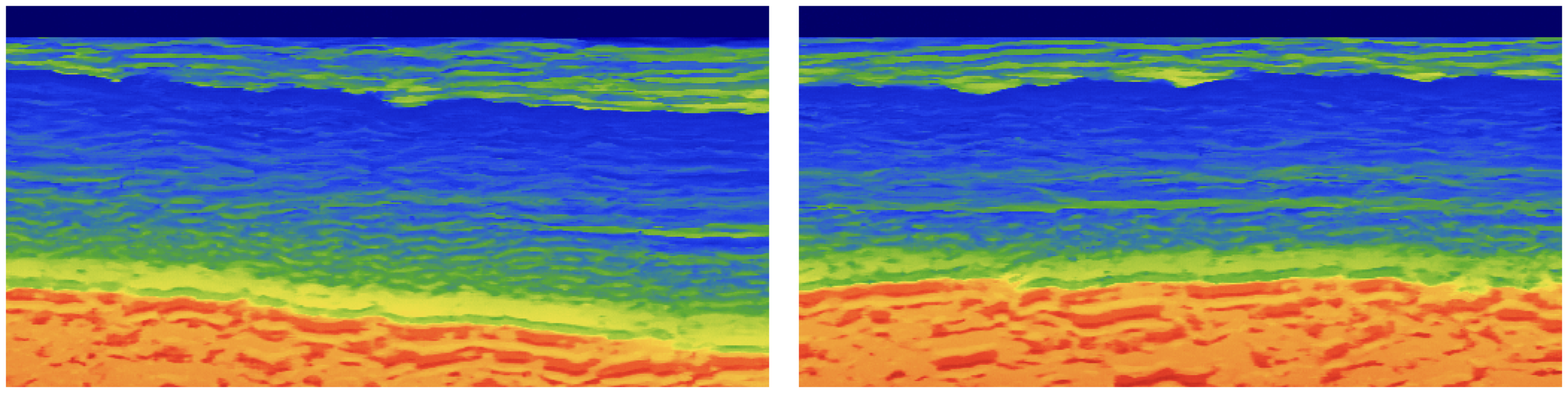}

}

\caption{\label{fig-gensamples}Generative samples from our trained
prior. Their similarity to the training samples in
Figure~\ref{fig-compasssamples} suggests that our normalizing flow has
learned a useful prior.}

\end{figure}

To illustrate the practical use of NFs as priors in seismic inverse
problems, we trained an NF on slices from the Compass model (Jones et
al. 2012). In Figure~\ref{fig-gensamples}, we compare generative samples
from the NF with the slices used to train the model shown in
Figure~\ref{fig-compasssamples}. Although there are still
irregularities, the model has learned important qualitative aspects of
the model that will be useful in inverse problems. To demonstrate this
usefulness, we test our prior on an FWI inverse problem. Because our NF
prior is trained independently, it is flexible and can easily be plugged
into different inverse problems.

Our FWI experiment includes: ocean bottom nodes, Ricker wavelet with no
energy below 4Hz, additive colored Gaussian noise that has the same
bandwidth as the noise-free data. For FWI with our learned prior, we
minimize

\begin{equation}\protect\hypertarget{eq-fwi-prior}{}{
\underset{\mathbf{z}}{\operatorname{minimize}} \quad \frac{1}{2}\|\mathbf{F}(\mathcal{G}_{\theta^\ast}(\mathbf{z}))\mathbf{q} - \mathbf{d}\|^2_2 + \frac{\lambda}{2}\|\mathbf{z}\|^2_2,
}\label{eq-fwi-prior}\end{equation}

where \(\mathcal{G}_{\theta^\ast}\) is a pretrained NF with weights
\(\theta^\ast\). After training, the inverse of the NF maps realistic
Compass-like Earth samples to white noise---i.e.,
\(\mathcal{G}^{-1}_{\theta^\ast}(\mathbf{m}) = \mathbf{z} \sim \mathcal{N}(0, I)\).
Since the NFs are designed to be invertible, the action of the
pretrained NF, \({\mathcal{G}_{\theta^\ast}}\), on Gaussian noise
\(\mathbf{z}\) produces realistic samples of Earth models (see
Figure~\ref{fig-gensamples}). We use this capability in the above
equation where the unknown model parameters in \(\mathbf{m}\) are
reparameterized by \(\mathcal{G}_{\theta^\ast}(\mathbf{z})\). The
regularization term, \(\frac{\lambda}{2}\|\mathbf{z}\|^2_2\), penalizes
the latent variable \(\mathbf{z}\) with large \(\ell_2\) norm, where
\(\lambda\) balances the misfit and regularization terms. Consequently,
this learned regularizer encourages FWI results that are more likely to
be realistic Earth models (Asim et al. 2020). However, notice that the
optimization routine now requires differentiation through both the
physical operator (wave physics, \(\mathbf{F}\)) and the pretrained NF
(\(\mathcal{G}_{\theta^\ast}\)), and only a true invertible
implementation like ours, with minimal memory imprint for both training
and inference, can provide scalability.

Thanks to the \href{https://github.com/slimgroup/JUDI.jl}{JUDI.jl}'s
\texttt{rrule} for \texttt{F} and
\href{https://github.com/slimgroup/InvertibleNetworks.jl}{InvertibleNetworks.jl}'s
\texttt{rrule} for \texttt{G}, integration of machine learning with FWI
becomes straightforward involving replacement of \texttt{m} by
\texttt{G(z)} on line 6. Minimizing the objective function in
Equation~\ref{eq-fwi-prior} now translates to

\begin{Shaded}
\begin{Highlighting}[]
\CommentTok{\# Load the pretrained NF and weights.}
\NormalTok{G }\OperatorTok{=} \FunctionTok{NetworkGlow}\NormalTok{(nc, nc\_hidden, depth, nscales)}
\FunctionTok{set\_params!}\NormalTok{(G, theta)}
\CommentTok{\# Set up the ADAM optimizer.}
\NormalTok{opt }\OperatorTok{=} \FunctionTok{ADAM}\NormalTok{()}
\CommentTok{\# Define the reparameterized loss function including penalty term.}
\FunctionTok{loss}\NormalTok{(z) }\OperatorTok{=} \FloatTok{.5f0} \OperatorTok{*} \FunctionTok{norm}\NormalTok{(}\FunctionTok{F}\NormalTok{(}\FunctionTok{G}\NormalTok{(z)) }\OperatorTok{*}\NormalTok{ q }\OperatorTok{{-}}\NormalTok{ d)}\OperatorTok{\^{}}\FloatTok{2f0} \OperatorTok{+} \FloatTok{.5f0} \OperatorTok{*}\NormalTok{ lambda }\OperatorTok{*} \FunctionTok{norm}\NormalTok{(z)}\OperatorTok{\^{}}\FloatTok{2f0}
\CommentTok{\# ADAM iterations.}
\ControlFlowTok{for}\NormalTok{ it }\OperatorTok{=} \FloatTok{1}\OperatorTok{:}\NormalTok{maxiter}
\NormalTok{    g }\OperatorTok{=} \FunctionTok{gradient}\NormalTok{(loss, z)[}\FloatTok{1}\NormalTok{]      }\CommentTok{\# gradient computation with AD}
    \FunctionTok{update!}\NormalTok{(opt, z, g)            }\CommentTok{\# update z with ADAM}
\ControlFlowTok{end}
\CommentTok{\# Convert latent variable to squared slowness.}
\NormalTok{m }\OperatorTok{=} \FunctionTok{G}\NormalTok{(z)}
\end{Highlighting}
\end{Shaded}

\begin{figure}

\begin{minipage}[t]{\linewidth}

{\centering 

\raisebox{-\height}{

\includegraphics[width=15cm]{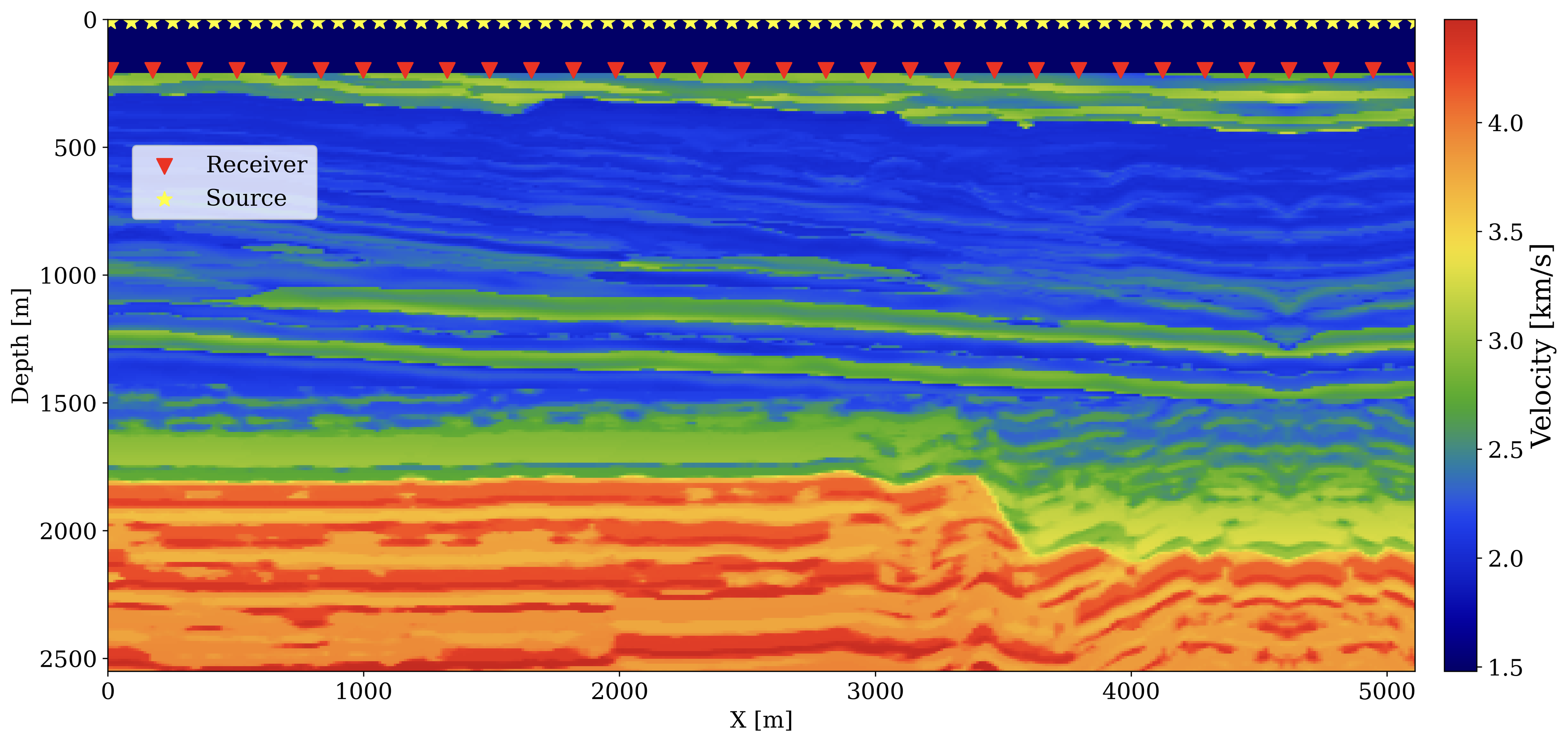}

}

}

\subcaption{\label{fig-fwi-true}}
\end{minipage}%
\newline
\begin{minipage}[t]{\linewidth}

{\centering 

\raisebox{-\height}{

\includegraphics[width=15cm]{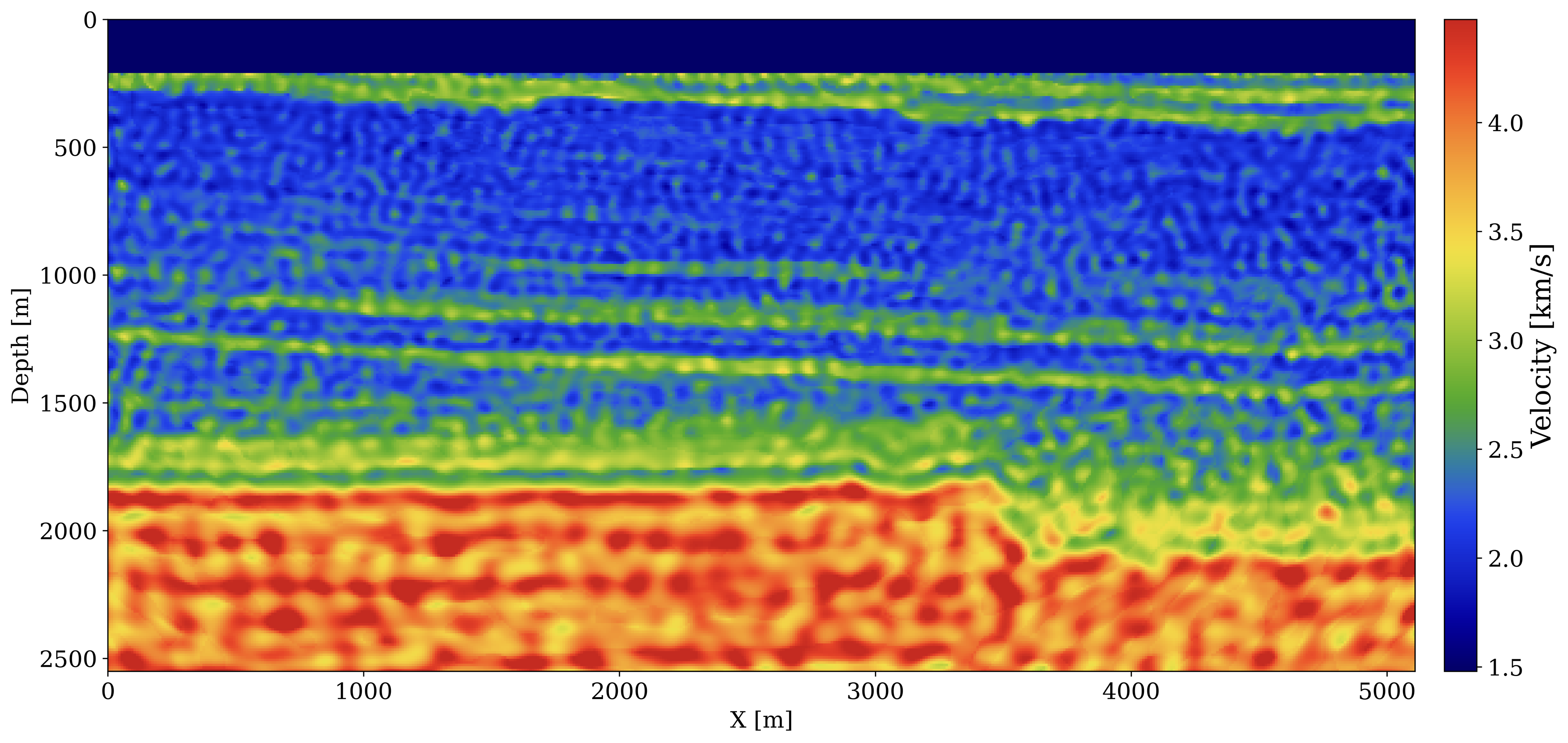}

}

}

\subcaption{\label{fig-fwi-vanilla}}
\end{minipage}%
\newline
\begin{minipage}[t]{\linewidth}

{\centering 

\raisebox{-\height}{

\includegraphics[width=15cm]{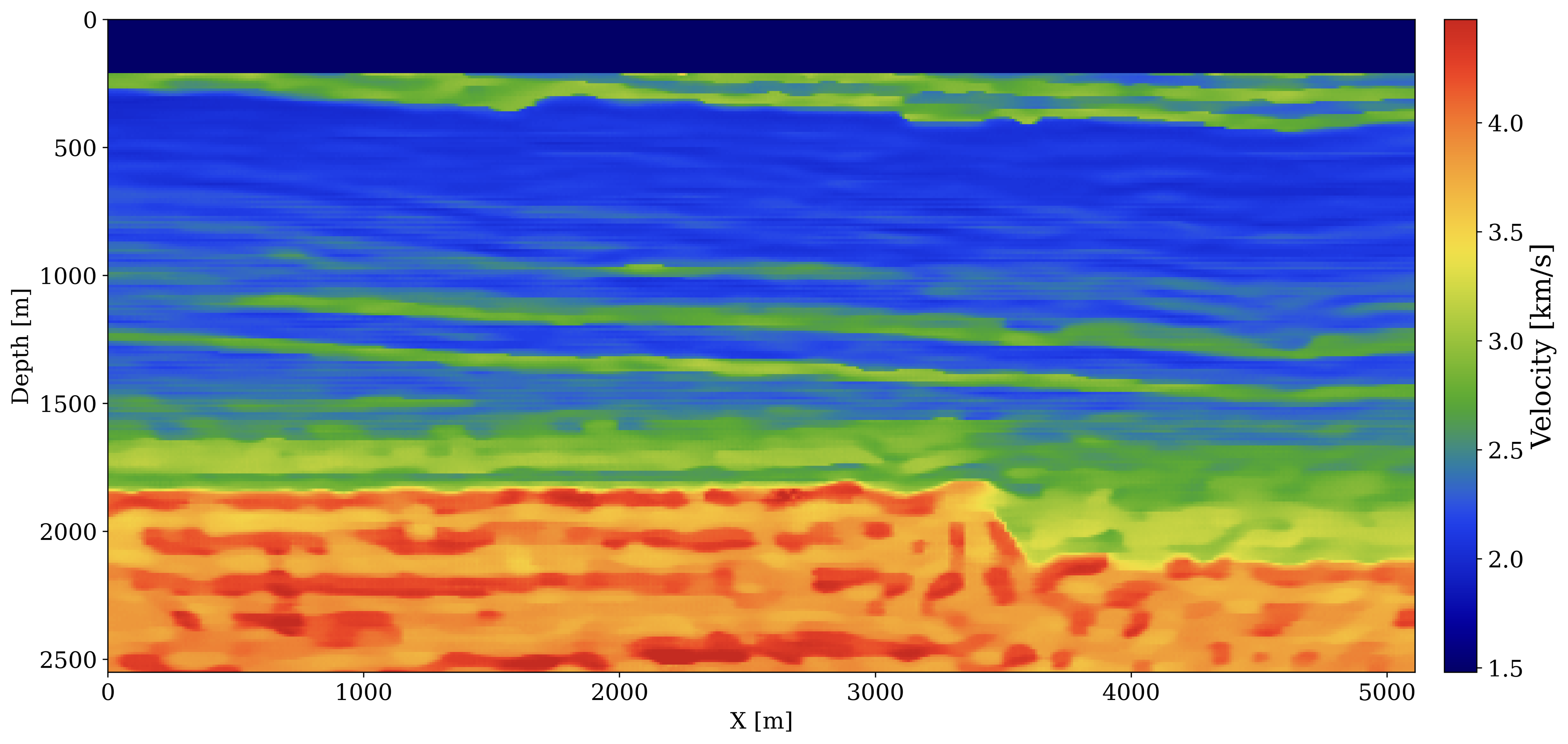}

}

}

\subcaption{\label{fig-fwi-nf}}
\end{minipage}%

\caption{\label{fig-fwideepprior}Results from using our normalizing flow
learned prior in FWI. \emph{(a)} Ground truth. \emph{(b)} Traditional
FWI without prior. \emph{(c)} Our FWI result with learned prior.}

\end{figure}

In Figure~\ref{fig-fwideepprior}, we compare the results of FWI with our
learned prior against unregularized FWI. Since our prior regularizes the
solution towards realistic models, we obtain a velocity estimate that is
closer to the ground truth.

Through this simple example, we demonstrated the ability to easily
integrate our state-of-the-art wave-equation propagators with the
Julia's differentiable programming system. By applying these design
principles to other components of the end-to-end inversion, we design a
seismic monitoring framework for real-world applications in subsurface
reservoirs.

\hypertarget{fluid-flow-simulation-and-permeability-inversion}{%
\subsubsection{Fluid-flow simulation and permeability
inversion}\label{fluid-flow-simulation-and-permeability-inversion}}

As stated earlier, our goal is to estimate the permeability from
time-lapse crosswell monitoring data collected at a CO\textsubscript{2}
injection site (cf. Figure~\ref{fig-end2end-acquisition}). Compared to
conventional seismic imaging, time-lapse monitoring of geological
storage differs because it aims to image time-lapse changes in the
CO\textsubscript{2} plume while obtaining estimates for the reservoir's
fluid-flow properties. This involves coupling wave modeling operators to
fluid-flow physics to track the CO\textsubscript{2} plumes underground.
The fluid-flow physics models the slow process of CO\textsubscript{2}
partly replacing brine in the pore space of the reservoir, which
involves solving the multiphase flow equations. For this purpose, we
need access to reservoir simulation software capable of modeling
two-phase (brine/CO\textsubscript{2}) flow. While a number of
proprietary and open-source reservoir simulators exist, including MRST
(Lie and Møyner 2021), GEOSX (Settgast et al. 2022) and Open Porous
Media (OPM) (Rasmussen et al. 2021), few support differentiation of the
simulator's output (CO\textsubscript{2} saturation) with respect to its
input (the spatial permeability distribution \(\mathbf{K}\) in
Figure~\ref{fig-end2end-latent}). We use the recently developed external
Julia package
\href{https://github.com/sintefmath/JutulDarcy.jl}{JutulDarcy.jl} that
supports Darcy flow and serves as a front-end to
\href{https://github.com/sintefmath/Jutul.jl}{Jutul.jl} (Møyner et al.
2023), which provides accurate Jacobians with respect to \(\mathbf{K}\).
\href{https://github.com/sintefmath/Jutul.jl}{Jutul.jl} is an implicit
solver for finite-volume discretizations that internally uses AD to
calculate the Jacobian. It has a performance and feature set comparable
to commercial multiphase flow simulators and accounts for realistic
effects (e.g., dissolution, inter-phase mass exchange, compressibility,
capillary effects) and residual trapping mechanisms. It also provides
accurate sensitivities through an adjoint formulation of the subsurface
multiphase flow equations. To integrate the Jacobian of this software
package into Julia's differentiable programming system, we wrote the
light ``wrapper package''
\href{https://github.com/slimgroup/JutulDarcyRules.jl}{JutulDarcyRules.jl}
(Yin and Louboutin 2023) that adds an \texttt{rrule} for the nonlinear
operator \(\mathcal{S}(\mathbf{K})\), which maps the permeability
distribution, \(\mathbf{K}\), to the spatially-varying
CO\textsubscript{2} concentration snapshots,
\(\mathbf{c}=\{\mathbf{c}^{i}\}_{i=1}^{n_v}\), over \(n_v\) monitoring
time-steps (cf. Figure~\ref{fig-end2end-latent}). Addition of this
\texttt{rrule} allows these packages to interoperate with other packages
in Julia's AD ecosystem. Below, we show a basic example where ADAM
algorithm is used to invert for subsurface permeability given the full
history of CO\textsubscript{2} concentration snapshots:

\begin{Shaded}
\begin{Highlighting}[]
\CommentTok{\# Generate CO2 concentration.}
\NormalTok{c }\OperatorTok{=} \FunctionTok{S}\NormalTok{(K\_true)}
\CommentTok{\# Set up ADAM optimizer.}
\NormalTok{opt }\OperatorTok{=} \FunctionTok{ADAM}\NormalTok{()}
\CommentTok{\# Define the loss function.}
\FunctionTok{loss}\NormalTok{(K) }\OperatorTok{=} \FloatTok{.5f0} \OperatorTok{*} \FunctionTok{norm}\NormalTok{(}\FunctionTok{S}\NormalTok{(K) }\OperatorTok{{-}}\NormalTok{ c)}\OperatorTok{\^{}}\FloatTok{2f0}
\CommentTok{\# ADAM iterations.}
\ControlFlowTok{for}\NormalTok{ it }\OperatorTok{=} \FloatTok{1}\OperatorTok{:}\NormalTok{maxiter}
\NormalTok{    g }\OperatorTok{=} \FunctionTok{gradient}\NormalTok{(loss, K)[}\FloatTok{1}\NormalTok{]      }\CommentTok{\# gradient computed with AD}
    \FunctionTok{update!}\NormalTok{(opt, K, g)            }\CommentTok{\# update K with ADAM}
\ControlFlowTok{end}
\end{Highlighting}
\end{Shaded}

During each iteration of the loop above, Julia's machine learning
package \href{https://github.com/FluxML/Flux.jl}{Flux.jl} (Michael Innes
et al. 2018; Mike Innes 2018) uses the custom gradient defined by the
aforementioned \texttt{rrule}, calling the high-performance adjoint code
from \href{https://github.com/sintefmath/JutulDarcy.jl}{JutulDarcy.jl}.
Our adaptable software framework also facilitates effortless
substitution of deep learning models in lieu of the numerical fluid-flow
simulator. In the next section, we introduce distributed Fourier neural
operators (\href{https://github.com/slimgroup/dfno}{dfno}) and discuss
how this neural surrogate contributes to our inversion framework.

\hypertarget{fourier-neural-operator-surrogates}{%
\subsubsection{Fourier neural operator
surrogates}\label{fourier-neural-operator-surrogates}}

While the integration of multiphase flow modeling into Julia
differentiable programming ecosystem opens the way to carry out
end-to-end inversions (as explained below), fluid-flow simulations are
computationally expensive---a notion compounded by the fact that these
simulations have to be done many times during inversion. For this
reason, we switch to a data-driven approach where a neural operator is
first trained on simulation examples, pairs
\(\{\mathbf{K},\, \mathcal{S}(\mathbf{K})\}\), to learn the mapping from
permeability models, \(\mathbf{K}\), to the corresponding
CO\textsubscript{2} snapshots,
\(\mathbf{c} := \mathcal{S}(\mathbf{K})\). After incurring initial
offline training costs, this neural surrogate provides a fast
alternative to numerical solvers with acceptable accuracy. Fourier
neural operators (FNOs, Z. Li et al. 2020), a recently introduced neural
network architecture, have been used successfully to simulate two-phase
flow during geological CO\textsubscript{2} storage projects (Wen et al.
2022). Independently, Yin et al. (2022) used a trained FNO to replace
the fluid-flow simulations as part of end-to-end inversion and showed
that AD of Julia's machine learning package can be used to compute
gradients with respect to the permeability using
\href{https://github.com/FluxML/Flux.jl}{Flux.jl}'s reverse-mode AD
system \href{https://github.com/FluxML/Zygote.jl}{Zygote.jl} (Michael
Innes 2018). After training, the above permeability inversion from
concentration snapshots, \(\mathbf{c}\), is carried out by simply
replacing \(\mathcal{S}\) by \(\mathcal{S}_{\mathbf{w}^\ast}\) with
\(\mathbf{w}^\ast\) being the weights of the pretrained FNO. Thanks to
the AD system, the gradient with respect to \(\mathbf{K}\) is computed
automatically. Thus, after loading the trained FNO and redefining the
operator \texttt{S}, the above code remains exactly the same. For
implementation details on the FNO and its training, we refer to Yin et
al. (2022) and Grady et al. (2022).

\hypertarget{putting-it-all-together}{%
\subsection{Putting it all together}\label{putting-it-all-together}}

As a final step in our end-to-end permeability inversion, we introduce a
nonlinear rock physics model, denoted by \(\mathcal{R}\). Based on the
patchy saturation model (Avseth, Mukerji, and Mavko 2010), this model
nonlinearly maps the time-lapse CO\textsubscript{2} saturations to
decreases in the seismic properties (compressional wavespeeds,
\(\mathbf{v}=\{\mathbf{v}^{i}\}_{i=1}^{n_v}\)) within the reservoir with
the Julia code

\begin{Shaded}
\begin{Highlighting}[]
\CommentTok{\# Patchy saturation function.}
\CommentTok{\# Input: CO2 saturation, velocity, density, porosity.}
\CommentTok{\# Optional: bulk modulus of mineral, brine, CO2; density of CO2, brine.}
\CommentTok{\# Output: velocity, density.}
\KeywordTok{function} \FunctionTok{Patchy}\NormalTok{(sw, vp, rho, phi;}
\NormalTok{    bulk\_min}\OperatorTok{=}\FloatTok{36.6f9}\NormalTok{, bulk\_fl1}\OperatorTok{=}\FloatTok{2.735f9}\NormalTok{, bulk\_fl2}\OperatorTok{=}\FloatTok{0.125f9}\NormalTok{,}
\NormalTok{    rhow}\OperatorTok{=}\FloatTok{7f2}\NormalTok{, rhoo}\OperatorTok{=}\FloatTok{1f3}\NormalTok{) }\KeywordTok{where}\NormalTok{ T}
    \CommentTok{\# Relate vp to vs, set modulus properties.}
\NormalTok{    vs }\OperatorTok{=}\NormalTok{ vp }\OperatorTok{./} \FunctionTok{sqrt}\NormalTok{(}\FloatTok{3f0}\NormalTok{)}
\NormalTok{    bulk\_sat1 }\OperatorTok{=}\NormalTok{ rho }\OperatorTok{.*}\NormalTok{ (vp}\OperatorTok{.\^{}}\FloatTok{2f0} \OperatorTok{.{-}} \FloatTok{4f0}\OperatorTok{/}\FloatTok{3f0} \OperatorTok{.*}\NormalTok{ vs}\OperatorTok{.\^{}}\FloatTok{2f0}\NormalTok{)}
\NormalTok{    shear\_sat1 }\OperatorTok{=}\NormalTok{ rho }\OperatorTok{.*}\NormalTok{ (vs}\OperatorTok{.\^{}}\FloatTok{2f0}\NormalTok{)}
    \CommentTok{\# Calculate bulk modulus if filled with 100\% CO2.}
\NormalTok{    patch\_temp }\OperatorTok{=}\NormalTok{ bulk\_sat1 }\OperatorTok{./}\NormalTok{ (bulk\_min }\OperatorTok{.{-}}\NormalTok{ bulk\_sat1)}
        \OperatorTok{.{-}}\NormalTok{ bulk\_fl1 }\OperatorTok{./}\NormalTok{ phi }\OperatorTok{./}\NormalTok{ (bulk\_min }\OperatorTok{.{-}}\NormalTok{ bulk\_fl1)}
        \OperatorTok{.+}\NormalTok{ bulk\_fl2 }\OperatorTok{./}\NormalTok{ phi }\OperatorTok{./}\NormalTok{ (bulk\_min }\OperatorTok{.{-}}\NormalTok{ bulk\_fl2)}
\NormalTok{    bulk\_sat2 }\OperatorTok{=}\NormalTok{ bulk\_min }\OperatorTok{./}\NormalTok{ (}\FloatTok{1f0} \OperatorTok{./}\NormalTok{ patch\_temp }\OperatorTok{.+} \FloatTok{1f0}\NormalTok{)}
    \CommentTok{\# Calculate new bulk modulus as weighted harmonic average.}
\NormalTok{    bulk\_new }\OperatorTok{=} \FloatTok{1f0} \OperatorTok{/}\NormalTok{ ((}\FloatTok{1f0} \OperatorTok{.{-}}\NormalTok{ sw) }\OperatorTok{./}\NormalTok{ (bulk\_sat1 }\OperatorTok{.+} \FloatTok{4f0}\OperatorTok{/}\FloatTok{3f0} \OperatorTok{*}\NormalTok{ shear\_sat1) }
    \OperatorTok{+}\NormalTok{ sw }\OperatorTok{./}\NormalTok{ (bulk\_sat2 }\OperatorTok{+} \FloatTok{4f0}\OperatorTok{/}\FloatTok{3f0} \OperatorTok{*}\NormalTok{ shear\_sat1)) }\OperatorTok{{-}} \FloatTok{4f0}\OperatorTok{/}\FloatTok{3f0} \OperatorTok{*}\NormalTok{ shear\_sat1}
    \CommentTok{\# Calculate new density and velocity.}
\NormalTok{    rho\_new }\OperatorTok{=}\NormalTok{ rho }\OperatorTok{+}\NormalTok{ phi }\OperatorTok{.*}\NormalTok{ sw }\OperatorTok{*}\NormalTok{ (rhow }\OperatorTok{{-}}\NormalTok{ rhoo)}
\NormalTok{    vp\_new }\OperatorTok{=} \FunctionTok{sqrt}\NormalTok{.((bulk\_new }\OperatorTok{.+} \FloatTok{4f0}\OperatorTok{/}\FloatTok{3f0} \OperatorTok{*}\NormalTok{ shear\_sat1) }\OperatorTok{./}\NormalTok{ rho\_new)}
    \ControlFlowTok{return}\NormalTok{ vp\_new, rho\_new}
\KeywordTok{end}
\end{Highlighting}
\end{Shaded}

We map the changes in the wavespeeds to time-lapse seismic data,
\(\mathbf{d}=\{\mathbf{d}^{i}\}_{i=1}^{n_v}\), via the blockdiagonal
seismic modeling\footnote{Note, we parameterized this forward modeling
  in terms of the compressional wavespeed.} operator
\(\mathcal{F}(\mathbf{v})=\operatorname{diag}\left(\left\{\mathbf{F}^{i}(\mathbf{v}^{i})\mathbf{q}^{i}\right\}_{i=1}^{n_v}\right)\).
In this formulation, the single vintage forward operators
\(\mathbf{F}^{i}\) and corresponding sources, \(\mathbf{q}^{i}\), are
allowed to vary between vintages.

With the fluid-flow (surrogate) solver, \(\mathcal{S}\), the rock
physics module, \(\mathcal{R}\), and wave physics module,
\(\mathcal{F}\), in place, along with regularization via
reparametrization using \(\mathcal{G}_{\theta^\ast}\), we are now in a
position to formulate the desired end-to-end inversion problem as

\begin{equation}\protect\hypertarget{eq-end2end}{}{
\underset{\mathbf{z}}{\operatorname{minimize}}\quad \frac{1}{2}\|\mathcal{F}\circ\mathcal{R}\circ\mathcal{S}\left(\mathcal{G}_{\theta^\ast}(\mathbf{z})\right)- \mathbf{d}\|^2_2 + \frac{\lambda}{2}\|\mathbf{z}\|^2_2, 
}\label{eq-end2end}\end{equation}

where the inverted permeability can be calculated by
\(\mathbf{K}^\ast=\mathcal{G}_{\theta^\ast}(\mathbf{z}^\ast)\) with
\(\mathbf{z}^\ast\) the latent space minimizer of
Equation~\ref{eq-end2end}. As illustrated in
Figure~\ref{fig-end2end-latent}, we obtain the nonlinear end-to-end map
by composing the fluid-flow, rock, and wave physics, according to
\(\mathcal{F}\circ\mathcal{R}\circ\mathcal{S}\). The corresponding Julia
code reads

\begin{Shaded}
\begin{Highlighting}[]
\CommentTok{\# Set up ADAM optimizer.}
\NormalTok{opt }\OperatorTok{=} \FunctionTok{ADAM}\NormalTok{()}
\CommentTok{\# Define the reparameterized loss function including penalty term.}
\FunctionTok{loss}\NormalTok{(z) }\OperatorTok{=} \FloatTok{.5f0} \OperatorTok{*} \FunctionTok{norm}\NormalTok{(}\FunctionTok{F}\NormalTok{(}\FunctionTok{R}\NormalTok{(}\FunctionTok{S}\NormalTok{(}\FunctionTok{G}\NormalTok{(z)))) }\OperatorTok{{-}}\NormalTok{ d)}\OperatorTok{\^{}}\FloatTok{2f0} \OperatorTok{+} \FloatTok{.5f0} \OperatorTok{*}\NormalTok{ lambda }\OperatorTok{*} \FunctionTok{norm}\NormalTok{(z)}\OperatorTok{\^{}}\FloatTok{2f0}
\CommentTok{\# ADAM iterations.}
\ControlFlowTok{for}\NormalTok{ it }\OperatorTok{=} \FloatTok{1}\OperatorTok{:}\NormalTok{maxiter}
\NormalTok{    g }\OperatorTok{=} \FunctionTok{gradient}\NormalTok{(loss, z)[}\FloatTok{1}\NormalTok{]      }\CommentTok{\# gradient computed by AD}
    \FunctionTok{update!}\NormalTok{(opt, z, g)            }\CommentTok{\# update z by ADAM}
\ControlFlowTok{end}
\CommentTok{\# Convert latent variable to permeability.}
\NormalTok{K }\OperatorTok{=} \FunctionTok{G}\NormalTok{(z)}
\end{Highlighting}
\end{Shaded}

This end-to-end inversion procedure, which utilizes a learned deep prior
and a pretrained FNO surrogate, was successfully employed by Yin et al.
(2022) on a simple stylistic blocky high-low permeability model. The
procedure involves using AD, with \texttt{rrule} for the wave and fluid
physics, in combination with innate AD capabilities to compute the
gradient of the objective in Equation~\ref{eq-end2end}, which
incorporates fluid-flow, rock, and wave physics. Below, we share early
results from applying the proposed end-to-end inversion in a more
realistic setting derived from real data (cf.
Figure~\ref{fig-end2end-acquisition}).

\hypertarget{preliminary-inversion-results}{%
\subsection{Preliminary inversion
results}\label{preliminary-inversion-results}}

\begin{figure}

\begin{minipage}[t]{0.50\linewidth}

{\centering 

\raisebox{-\height}{

\includegraphics{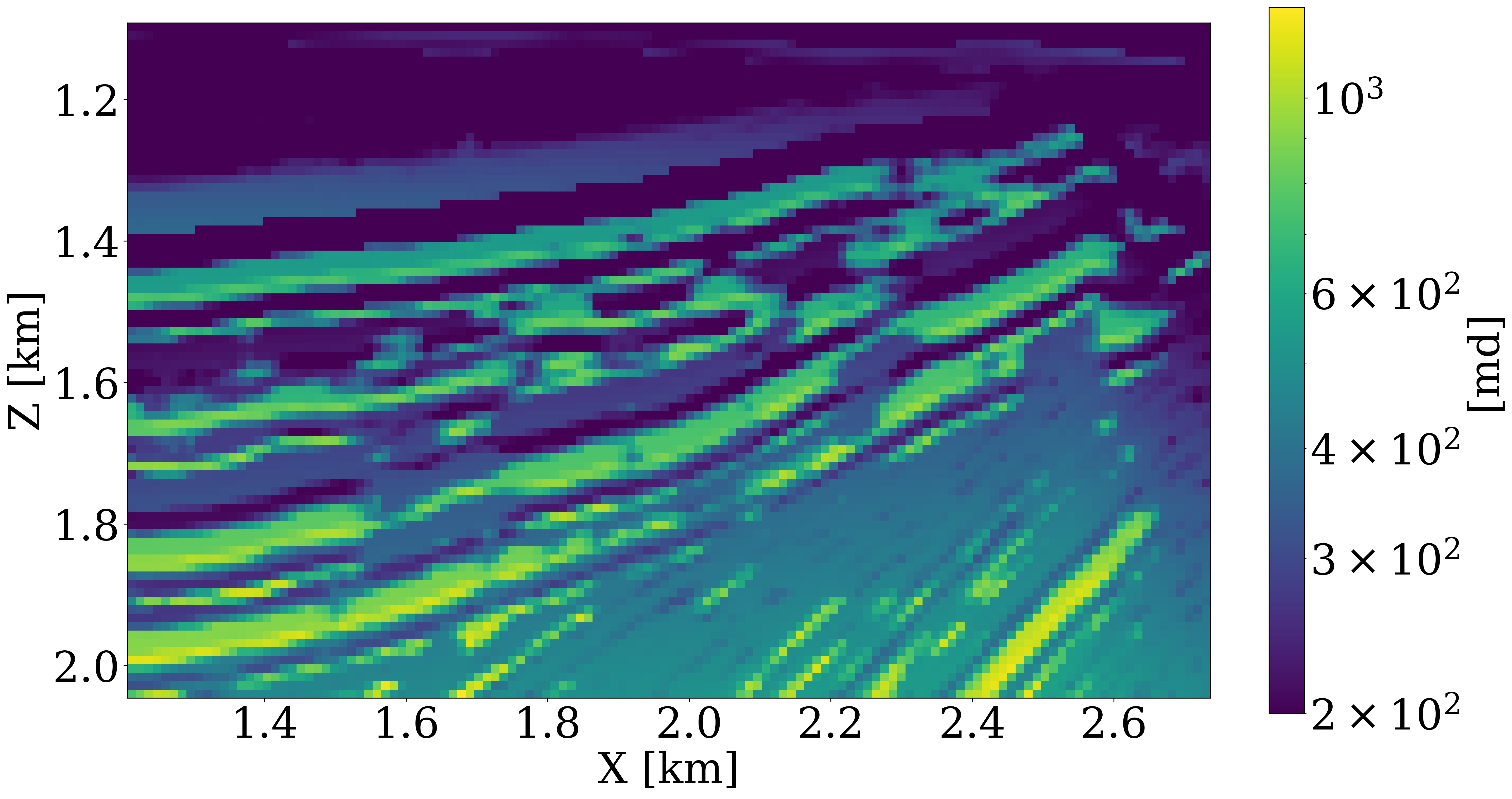}

}

}

\subcaption{\label{fig-inv-perm-K}}
\end{minipage}%
\begin{minipage}[t]{0.50\linewidth}

{\centering 

\raisebox{-\height}{

\includegraphics{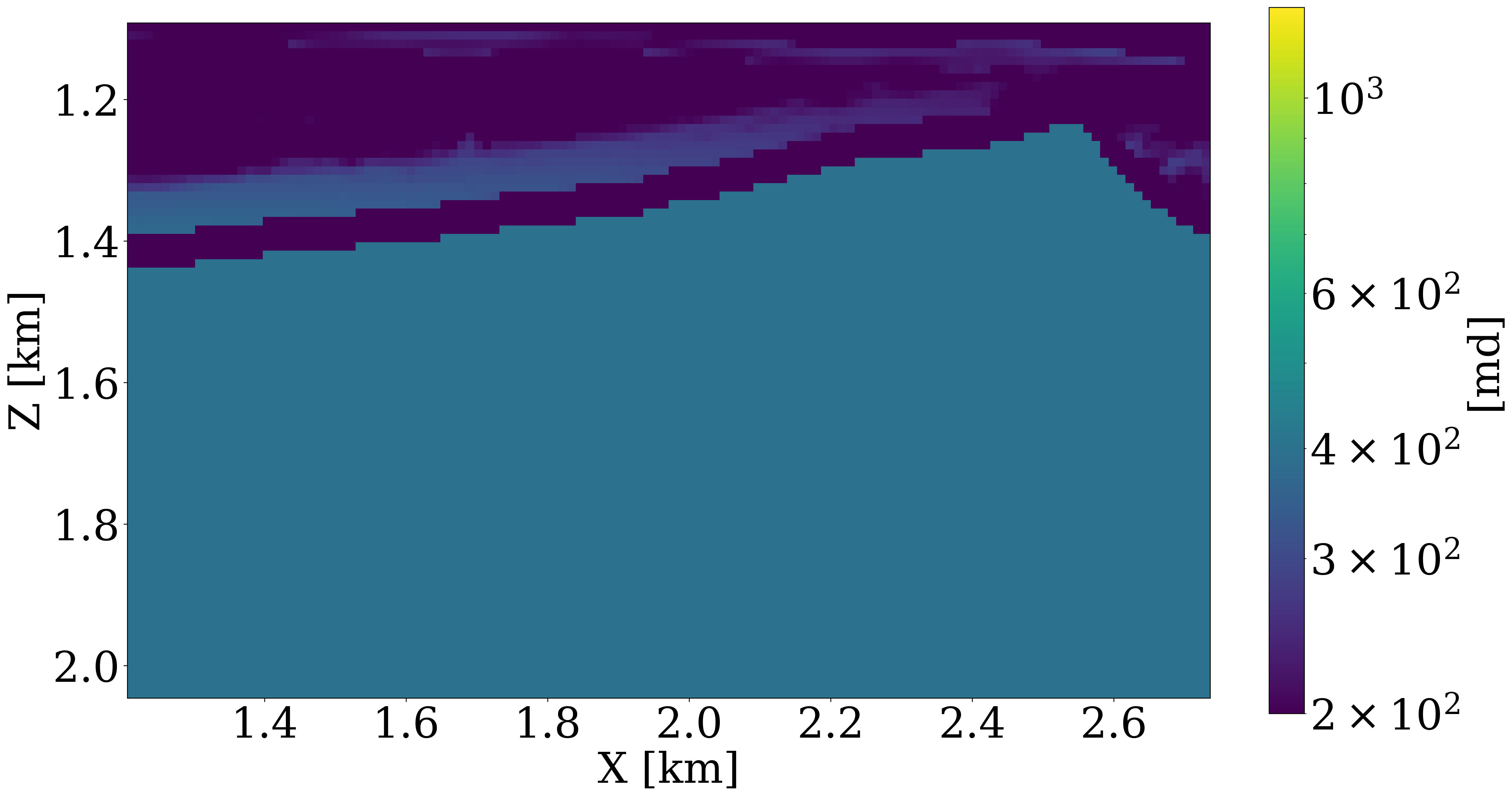}

}

}

\subcaption{\label{fig-inv-perm-K0}}
\end{minipage}%
\newline
\begin{minipage}[t]{0.50\linewidth}

{\centering 

\raisebox{-\height}{

\includegraphics{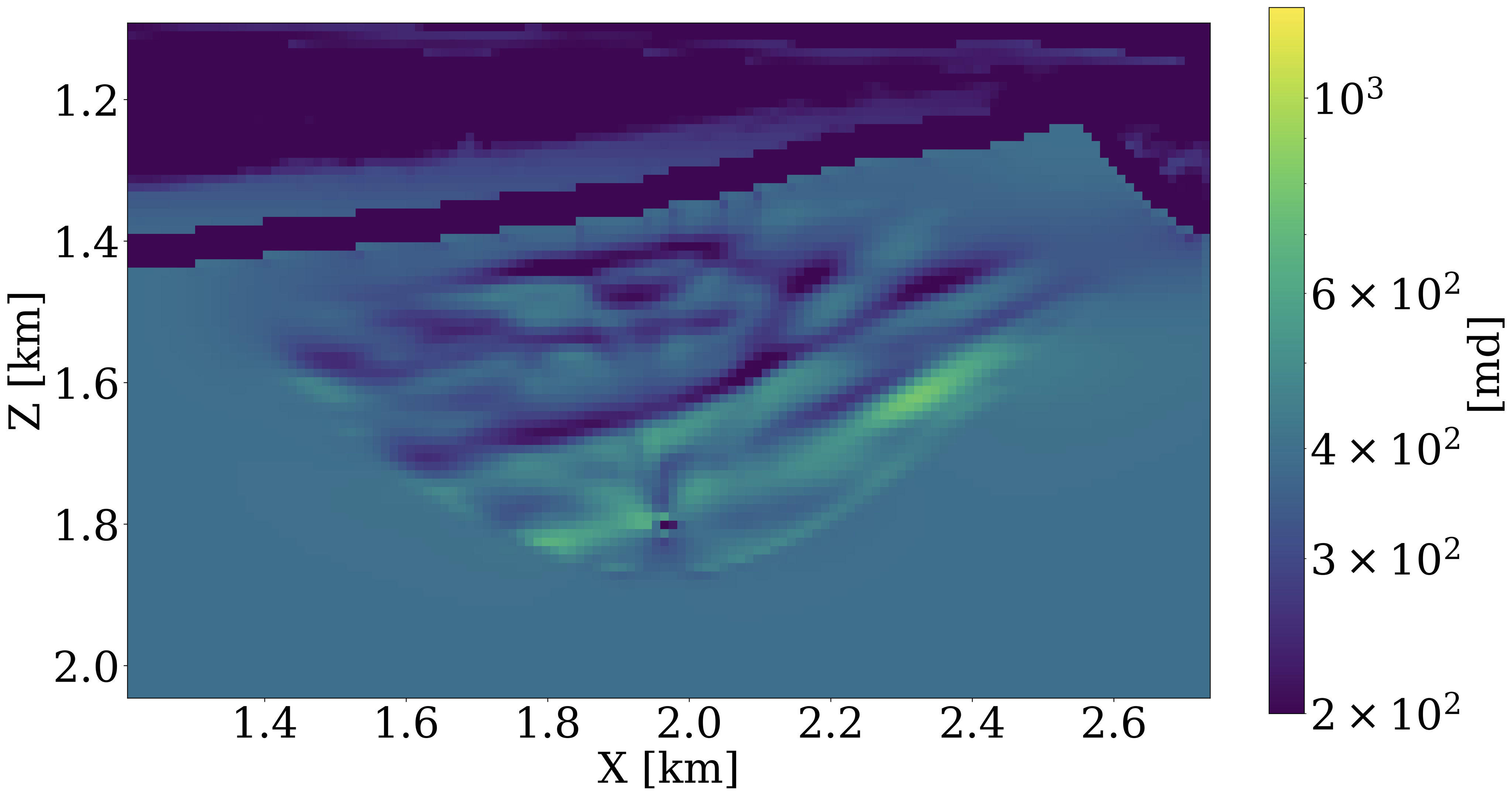}

}

}

\subcaption{\label{fig-inv-perm-K-inv}}
\end{minipage}%
\begin{minipage}[t]{0.50\linewidth}

{\centering 

\raisebox{-\height}{

\includegraphics{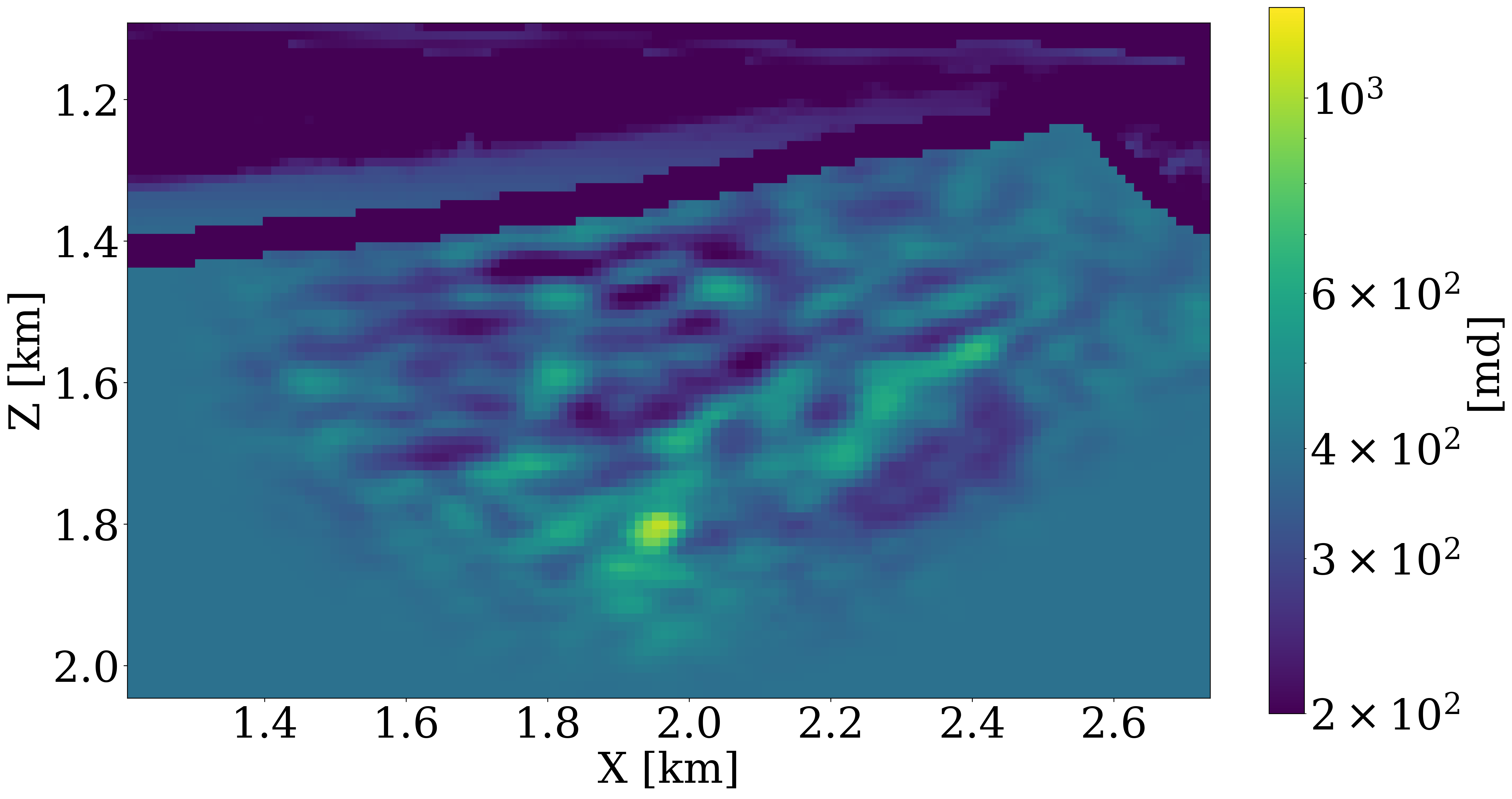}

}

}

\subcaption{\label{fig-inv-perm-K-invfno}}
\end{minipage}%

\caption{\label{fig-inv-perm}Fifteen-year time-lapse seismic end-to-end
permeability inversion with physics-based and surrogate fluid-flow
simulations. \emph{(a)} Ground truth permeability. \emph{(b)} Initial
permeability with homogeneous values in the reservoir. \emph{(c)}
Inverted permeability from physics-based inversion. \emph{(d)} Inverted
permeability with neural surrogate approximation.}

\end{figure}

\begin{figure}

{\centering 

\includegraphics{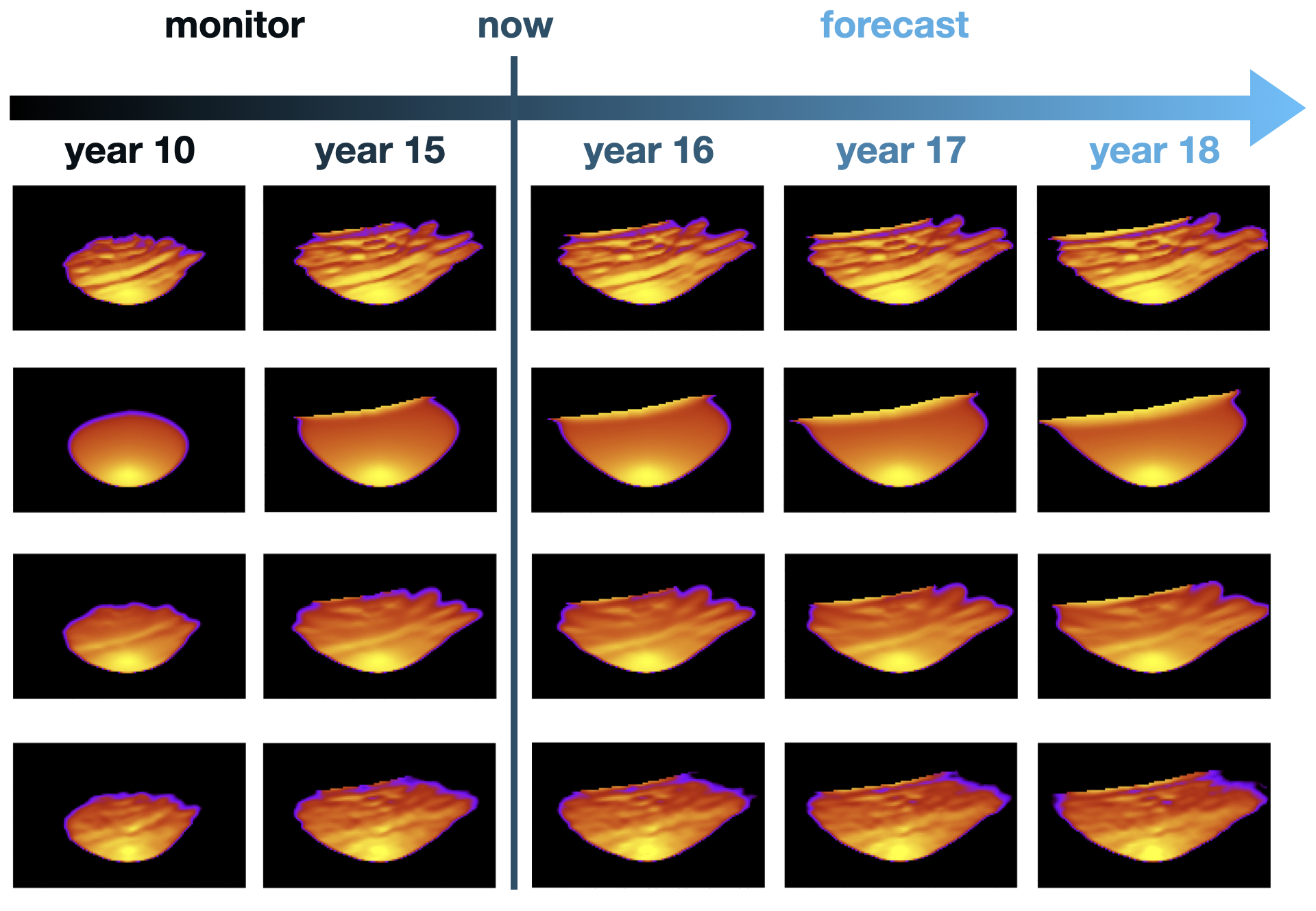}

}

\caption{\label{fig-pred-co2}CO\textsubscript{2} plume estimation and
prediction. The first two columns are the CO\textsubscript{2}
concentration snapshots at year 10 and year 15 of the first 15 years of
simulation monitored seismically. The last three columns are forecasted
snapshots at years 16, 17, 18, where no seismic data is available. First
row corresponds to the ground truth CO\textsubscript{2} plume simulated
by the unseen ground truth permeability model. Second row contains plume
simulations in the starting model. Rows three and four contain estimated
and predicted CO\textsubscript{2} plumes for the physics-based and
surrogate-based permeability inversion.}

\end{figure}

While initial results by Yin et al. (2022) were encouraging and showed
strong benefits from the learned prior, the permeability model and fluid
flow simulations considered in their study were too simplistic. To
evaluate the proposed end-to-end inversion methodology in a more
realistic setting, we consider the permeability model plotted in
Figure~\ref{fig-inv-perm-K}, which we derived from a slice of the
Compass model (Jones et al. 2012) shown in
Figure~\ref{fig-end2end-acquisition}. To generate realistic
CO\textsubscript{2} plumes in this model, we generate immiscible and
compressible two-phase flow simulations with
\href{https://github.com/sintefmath/JutulDarcy.jl}{JutulDarcy.jl} over a
period of 18 years with 5 snapshots plotted at years 10, 15, 16, 17, and
18. These CO\textsubscript{2} snapshots are shown in the first row of
Figure~\ref{fig-pred-co2}. Next, given the fluid-flow simulation, we use
the patchy saturation model (Avseth, Mukerji, and Mavko 2010) to convert
each CO\textsubscript{2} concentration snapshot,
\(\mathbf{c}^{i},\, i=1\ldots n_v\) to corresponding wavespeed model,
\(\mathbf{v}^{i},\, i=1\ldots n_v\) with
\(\mathbf{v}=\mathcal{R}(\mathbf{c})\). We then use
\href{https://github.com/slimgroup/JUDI.jl}{JUDI.jl} to generate
synthetic time-lapse data, \(\mathbf{d}^{i},\, i=1\ldots n_v\), for each
vintage.

During the inversion, the first 15 years of time-lapse data,
\(\mathbf{d}^{i},\, i=1\ldots 15\), from the above synthetic experiment
are inverted with permeabilities within the reservoir initialized by a
single reasonable value as shown in Figure~\ref{fig-inv-perm-K0}.
Inversion results obtained after 25 passes through the data for the
physics-driven two-phase flow solver and its learned neural surrogate
approximation are included in Figure~\ref{fig-inv-perm-K-inv} and
Figure~\ref{fig-inv-perm-K-invfno}, respectively. Both results were
obtained with 200 iterations of the code block shown above. Each
time-lapse vintage consist of 960 receivers and 32 shots. To limit the
number of wave-equation solves, gradients were calculated for only four
randomly selected shots with replacement per iteration. While these
results obtained without learned regularization are somewhat
preliminary, they lead to the following observations. First, both
inversion results for the permeability follow the inverted cone shape of
the CO\textsubscript{2}. This is to be expected because permeability can
only be inverted where CO\textsubscript{2} has flown over the first 15
years. Second, the inverted permeability follows trends of this strongly
heterogeneous model. Third, as expected details and continuity of the
results obtained with the two-phase flow solver are better. In part,
this can be explained by the fact that there are no guarantees that the
model iterations remain with the statistical distribution on which the
FNO was trained. Fourth, the implementation of this workflow greatly
benefited from the software design principles listed above. For
instance, the use of abstractions made it trivial to replace
physics-driven two-phase flow solvers with their learned counterparts.

Despite the inversion results being preliminary, the 18 year
CO\textsubscript{2} simulations in both inverted permeability models are
reasonable when comparing the true plume development plotted in the top
row of Figure~\ref{fig-pred-co2} with plumes simulated from the inverted
models plotted in rows three and four of Figure~\ref{fig-pred-co2}.
While certain details are missing in the estimates for the past,
current, and predicted CO\textsubscript{2} concentrations, the inversion
constitutes a considerable improvement compared to plumes generated in
the starting model for the permeability plotted in the second row of
Figure~\ref{fig-pred-co2}. An early version of the presented workflow
can be found in the Julia Package
\href{https://github.com/slimgroup/Seis4CCS.jl}{Seis4CCS.jl}. As the
project matures, updated workflows and codes will be pushed to GitHub.

\hypertarget{remaining-challenges}{%
\subsection{Remaining challenges}\label{remaining-challenges}}

So far, we hope we were able to convince the reader that working with
abstractions certainly has its benefits. Thanks to the math-inspired
abstractions, which naturally lead to modularity and separation of
concerns, we were able to accelerate the research \& development cycle
for the end-to-end inversion. As a result, we created a development
environment that allowed us to include machine learning techniques.
Relatively late in the development cycle, it also gave us the
opportunity to swap out the original 2D reservoir simulation code for a
much more powerful and fully-featured industry-strength 3D code
developed by a national lab. What we unfortunately not yet have been
able to do is to demonstrate our ability to scale this end-to-end
inversion to 3D, while both the
\href{https://github.com/devitocodes/devito}{Devito}-based propagators
and \href{https://github.com/sintefmath/Jutul.jl}{Jutul.jl}'s fluid-flow
simulations both have been demonstrated on industry-scale problems.
Unfortunately, lack of access to large-scale computational resources
makes it challenging in an academic environment to validate the proposed
methodology on 4D synthetic and field data even though the computational
toolchain presented in this paper is fully differentiable and in
principle capable of scale-up. Most components have been separately
tested and verified on realistic 3D examples (Grady et al. 2022; Møyner
and Bruer 2023; Mathias Louboutin and Herrmann 2022; Mathias Louboutin
and Herrmann 2023) and efforts are underway to remove fundamental memory
and other bottlenecks.

\hypertarget{scale-up-normalizing-flows}{%
\subsubsection{Scale-up normalizing
flows}\label{scale-up-normalizing-flows}}

Generative models, and NFs included, call for relatively large training
sets and large computational resources for training. While efforts have
been made to create training sets for the more traditional machine
learning tasks, no public-domain training set exists that contains
realistic 3D examples. The good news is that normalizing flows (Rezende
and Mohamed 2015) have a small memory footprint compared to diffusion
models (Song et al. 2020), so training this type of network will be
feasible when training sets and compute become available. In our
laboratory, we were already able to successfully train and evaluate NFs
on \(256 \times 256 \times 64\) models.

\hypertarget{scale-up-neural-operators}{%
\subsubsection{Scale-up neural
operators}\label{scale-up-neural-operators}}

Since the seminal paper by (Z. Li et al. 2020), there has been a flurry
of publications on the use of FNOs as neural surrogates for expensive
multiphase fluid-flow solvers used to simulate CO\textsubscript{2}
injection as part of geological storage projects (Wen et al. 2022,
2023). While there is good reason for this excitement, challenges remain
when scaling this technique to realistic 3D problems. In that case,
additional measures have to be taken. For instance, by nesting FNOs Wen
et al. (2023) were able to divide 3D domains into smaller hierarchical
subdomains centered around the wells, an approach that is only viable
when certain assumptions are met. Because of this nested decomposition,
these authors avoid the large memory footprint of 3D FNOs and report
many orders of magnitude speedup. Given the potential impact of
irregular CO\textsubscript{2} flow, e.g., leakage, we as much as
possible try to avoid making assumptions on the flow behavior and
propose an accurate distributed Fourier neural operator
(\href{https://github.com/slimgroup/dfno}{dfno}) structure based on a
domain decomposition of the network's input and network weights (Grady
et al. 2022). By using DistDL (Hewett and Grady II 2020), a software
package that supports ``model parallelism'' in machine learning, our
\href{https://github.com/slimgroup/dfno}{dfno} partitions the input data
and network weights across multiple GPUs such that each partition is
able to fit in the memory of a single GPU. As reported by Grady et al.
(2022), our work demonstrated validity of
\href{https://github.com/slimgroup/dfno}{dfno} on realistic problem and
reasonable training set (permeability/CO\textsubscript{2} concentration
pairs) sizes, for permeability models derived from the Sleipner
benchmark model (Furre et al. 2017). On 16 timesteps and models of size
\(64 \times 118 \times 263\), we reported from our perspective a more
realistic speedup of over \(1300\times\) compared to the simulation time
on Open Porous Media (Rasmussen et al. 2021), one of the leading
open-source reservoir simulators. These results confirm a similar
indepedent approach advocated by P. A. Witte et al. (2022). Even though
we are working with our industrial partners and Extreme Scale Solutions
to further improve these numbers, we are confident that distributed FNOs
are able to scale to 3D with a high degree of parallel efficiency.

\hypertarget{towards-scalable-open-source-software}{%
\subsubsection{Towards scalable open-source
software}\label{towards-scalable-open-source-software}}

In addition to allowing for reproduction of published results, we are
big advocates of pushing out scalable open-source software to help with
the energy transition and with combating climate change. As observed in
other fields, most notably in machine learning, open-source software
leads to accelerated rates of innovation, a feature we need as an
industry faced with major challenges. Despite the above exposition on
our experiences implementing end-to-end permeability inversion, this
work constitutes a snapshot of an ongoing project. However, many of the
software components listed in Table~\ref{tbl-softwares} are in an
advanced stage of development and to a large degree ready to be tested
in 3D and ultimately on field data. For instance, all our software
supports large-scale 3D simulation and AD. In addition, we are in
advanced state of development to support GPU for all codes. For those
curious on future developments, we also include below the Julia package
\href{https://github.com/slimgroup/ParametricOperators.jl}{ParametricOperators.jl},
which is designed to allow for high-dimensional parallel tensor
manipulations in support of future Julia-native implementations of
distributed FNOs.

The work presented in this paper would not have been possible without
open-source efforts from other groups, most notably by researchers at
the UK's Imperial College London who spearheaded the development of
\href{https://github.com/devitocodes/devito}{Devito} and researchers at
Norway's SINTEF. By integrating these packages into
\href{https://julialang.org}{Julia}'s agile differentiable programming
environment, we believe that we are well on our way to arrive at a
software environment that is much more viable than the sum of its parts.
We welcome readers to check \url{https://github.com/slimgroup} for the
latest developments.

\hypertarget{tbl-softwares}{}
\begin{longtable}[]{@{}
  >{\raggedright\arraybackslash}p{(\columnwidth - 8\tabcolsep) * \real{0.3580}}
  >{\raggedright\arraybackslash}p{(\columnwidth - 8\tabcolsep) * \real{0.0741}}
  >{\raggedright\arraybackslash}p{(\columnwidth - 8\tabcolsep) * \real{0.0741}}
  >{\raggedright\arraybackslash}p{(\columnwidth - 8\tabcolsep) * \real{0.0741}}
  >{\raggedright\arraybackslash}p{(\columnwidth - 8\tabcolsep) * \real{0.4198}}@{}}
\caption{\label{tbl-softwares}Current state of
\href{https://github.com/slimgroup}{SLIM}'s software stack. To underline
collaboration and active participation in other open-source projects, we
included the external software packages (denoted by \(^*\)) as well as
how these are integrated into our software framework.}\tabularnewline
\toprule()
\begin{minipage}[b]{\linewidth}\raggedright
Package
\end{minipage} & \begin{minipage}[b]{\linewidth}\raggedright
3D
\end{minipage} & \begin{minipage}[b]{\linewidth}\raggedright
GPU
\end{minipage} & \begin{minipage}[b]{\linewidth}\raggedright
AD
\end{minipage} & \begin{minipage}[b]{\linewidth}\raggedright
Parallelism
\end{minipage} \\
\midrule()
\endfirsthead
\toprule()
\begin{minipage}[b]{\linewidth}\raggedright
Package
\end{minipage} & \begin{minipage}[b]{\linewidth}\raggedright
3D
\end{minipage} & \begin{minipage}[b]{\linewidth}\raggedright
GPU
\end{minipage} & \begin{minipage}[b]{\linewidth}\raggedright
AD
\end{minipage} & \begin{minipage}[b]{\linewidth}\raggedright
Parallelism
\end{minipage} \\
\midrule()
\endhead
\href{https://github.com/devitocodes/devito}{Devito}\(^*\) & yes & yes &
no & domain-decomposition via MPI, multi-threading via OpenMP \\
\href{https://github.com/slimgroup/JUDI.jl}{JUDI.jl} & yes & yes & yes &
multi-threading via OpenMP, task parallel \\
\href{https://github.com/slimgroup/JUDI4Cloud.jl}{JUDI4Cloud.jl} & yes &
yes & yes & multi-threading via OpenMP, task parallel \\
\href{https://github.com/slimgroup/InvertibleNetworks.jl}{InvertibleNetworks.jl}
& yes & yes & yes & Julia-native multi-threading \\
\href{https://github.com/slimgroup/dfno}{dfno} & yes & yes & yes &
domain-decomposition via MPI \\
\href{https://github.com/sintefmath/Jutul.jl}{Jutul.jl}\(^*\) & yes &
soon & yes & Julia-native multi-threading \\
\href{https://github.com/slimgroup/JutulDarcyRules.jl}{JutulDarcyRules.jl}
& yes & soon & yes & Julia-native multi-threading \\
\href{https://github.com/slimgroup/Seis4CCS.jl}{Seis4CCS.jl} & yes & yes
& yes & Julia-native multi-threading \\
\href{https://github.com/slimgroup/ParametricOperators.jl}{ParametricOperators.jl}
& yes & yes & yes & domain-decomposition via MPI, Julia-native
multi-threading \\
\bottomrule()
\end{longtable}

\hypertarget{conclusions}{%
\subsection{Conclusions}\label{conclusions}}

In this work, we introduced a software framework for geophysical inverse
problems and machine learning that provides a scalable, portable, and
interoperable environment for research and development at scale. We
showed that through carefully chosen design principles, software with
math-inspired abstractions can be created that naturally leads to
desired modularity and separation of concerns without sacrificing
performance. We achieve this by combining
\href{https://github.com/devitocodes/devito}{Devito}'s automatic code
generation for wave propagators with Julia's modern highly performant
and scalable programming capabilities, including differentiable
programming. Thanks to these features, we were able to quickly implement
a prototype, in principle scalable to 3D, for permeability inversion
from time-lapse crosswell seismic data. Aside from the use of proper
abstractions, our approach to solving this relatively complex
multiphysics problem extensively relied on Julia's innate algorithmic
differentiation capabilities, supplemented by auxiliary performant
derivatives for the wave/fluid-flow physics, and for components of the
machine learning. On account of these design choices, we developed an
agile and relatively easy to maintain compact software stack where
low-level code is hidden through a combination of math-inspired
abstractions, modern programming practices, and automatic code
generation.

\hypertarget{code-and-data-availability}{%
\subsection{Code and data
availability}\label{code-and-data-availability}}

Our software framework is organized into registered
\href{https://julialang.org}{Julia} packages, all of which can be found
on the \href{https://github.com/slimgroup}{SLIM} GitHub page
(\url{https://github.com/slimgroup}). The software packages described in
this paper are all open-source and released under the MIT license for
use by the community.

\hypertarget{acknowledgment}{%
\subsection{Acknowledgment}\label{acknowledgment}}

This research was carried out with the support of Georgia Research
Alliance and industrial partners of the ML4Seismic Center. The authors
thank Henryk Modzelewski (University of British Columbia) and Rishi Khan
(Extreme Scale Solutions) for constructive discussions. This work was
supported in part by the US National Science Foundation grant OAC
2203821 and the Department of Energy grant No.~DE-SC0021515.

\bibliography{biblio}
\hypertarget{references}{%
\subsubsection{References}\label{references}}

\hypertarget{refs}{}
\begin{CSLReferences}{1}{0}
\leavevmode\vadjust pre{\hypertarget{ref-fenics}{}}%
Alnaes, M. S., J. Blechta, J. Hake, A. Johansson, B. Kehlet, A. Logg, C.
Richardson, J. Ring, M. E. Rognes, and G. N. Wells. 2015. {``The
{FEniCS} Project Version 1.5.''} \emph{Archive of Numerical Software} 3.
\url{https://doi.org/10.11588/ans.2015.100.20553}.

\leavevmode\vadjust pre{\hypertarget{ref-pmlr-v119-asim20a}{}}%
Asim, Muhammad, Max Daniels, Oscar Leong, Ali Ahmed, and Paul Hand.
2020. {``Invertible Generative Models for Inverse Problems: Mitigating
Representation Error and Dataset Bias.''} In \emph{Proceedings of the
37th International Conference on Machine Learning}, 119:399--409.
Proceedings of Machine Learning Research. PMLR.
\url{http://proceedings.mlr.press/v119/asim20a.html}.

\leavevmode\vadjust pre{\hypertarget{ref-avseth2010quantitative}{}}%
Avseth, Per, Tapan Mukerji, and Gary Mavko. 2010. \emph{Quantitative
Seismic Interpretation: Applying Rock Physics Tools to Reduce
Interpretation Risk}. Cambridge university press.

\leavevmode\vadjust pre{\hypertarget{ref-friedlander2009SCAIMspot}{}}%
Berg, Ewout van den, and Michael P. Friedlander. 2009. {``Spot: A
Linear-Operator Toolbox for Matlab.''} University of British Columbia:
SCAIM Seminar; SCAIM Seminar.

\leavevmode\vadjust pre{\hypertarget{ref-Julia}{}}%
Bezanson, Jeff, Alan Edelman, Stefan Karpinski, and Viral B Shah. 2017.
{``Julia: A Fresh Approach to Numerical Computing.''} \emph{SIAM
{R}eview} 59 (1): 65--98. \url{https://doi.org/10.1137/141000671}.

\leavevmode\vadjust pre{\hypertarget{ref-dinh2016density}{}}%
Dinh, Laurent, Jascha Sohl-Dickstein, and Samy Bengio. 2016. {``{Density
estimation using Real NVP}.''} In \emph{{International Conference on
Learning Representations, {ICLR}}}.
\url{http://arxiv.org/abs/1605.08803}.

\leavevmode\vadjust pre{\hypertarget{ref-Madagascar}{}}%
Fomel, Sergey, Paul Sava, Ioan Vlad, Yang Liu, and Vladimir Bashkardin.
2013. {``Madagascar: Open-Source Software Project for Multidimensional
Data Analysis and Reproducible Computational Experiments.''}
\emph{Journal of Open Research Software} 1 (1).

\leavevmode\vadjust pre{\hypertarget{ref-FURRE20173916}{}}%
Furre, Anne-Kari, Ola Eiken, Håvard Alnes, Jonas Nesland Vevatne, and
Anders Fredrik Kiær. 2017. {``20 Years of Monitoring CO2-Injection at
Sleipner.''} \emph{Energy Procedia} 114: 3916--26.
https://doi.org/\url{https://doi.org/10.1016/j.egypro.2017.03.1523}.

\leavevmode\vadjust pre{\hypertarget{ref-dfno}{}}%
Grady, Thomas J., Infinoid, and Mathias Louboutin. 2022.
\emph{Slimgroup/Dfno: Optimal Comm} (version 0.3). Zenodo.
\url{https://doi.org/10.5281/zenodo.6981516}.

\leavevmode\vadjust pre{\hypertarget{ref-grady2022dfno}{}}%
Grady, Thomas J., Rishi Khan, Mathias Louboutin, Ziyi Yin, Philipp A.
Witte, Ranveer Chandra, Russell J. Hewett, and Felix J. Herrmann. 2022.
{``Model-Parallel Fourier Neural Operators as Learned Surrogates for
Large-Scale Parametric PDEs.''} \emph{CoRR}.
\url{https://doi.org/10.48550/ARXIV.2204.01205}.

\leavevmode\vadjust pre{\hypertarget{ref-herrmann2019NIPSliwcuc}{}}%
Herrmann, Felix J., Ali Siahkoohi, and Gabrio Rizzuti. 2019. {``Learned
Imaging with Constraints and Uncertainty Quantification.''}
\url{https://arxiv.org/pdf/1909.06473.pdf}.

\leavevmode\vadjust pre{\hypertarget{ref-hewett2020linear}{}}%
Hewett, Russell J, and Thomas J Grady II. 2020. {``A Linear Algebraic
Approach to Model Parallelism in Deep Learning.''} \emph{arXiv Preprint
arXiv:2006.03108}.

\leavevmode\vadjust pre{\hypertarget{ref-Zygote.jl-2018}{}}%
Innes, Michael. 2018. {``Don't Unroll Adjoint: Differentiating SSA-Form
Programs.''} \emph{CoRR} abs/1810.07951.
\url{http://arxiv.org/abs/1810.07951}.

\leavevmode\vadjust pre{\hypertarget{ref-Flux.jl-2018}{}}%
Innes, Michael, Elliot Saba, Keno Fischer, Dhairya Gandhi, Marco
Concetto Rudilosso, Neethu Mariya Joy, Tejan Karmali, Avik Pal, and
Viral Shah. 2018. {``Fashionable Modelling with Flux.''} \emph{CoRR}
abs/1811.01457. \url{https://arxiv.org/abs/1811.01457}.

\leavevmode\vadjust pre{\hypertarget{ref-innes:2018}{}}%
Innes, Mike. 2018. {``Flux: Elegant Machine Learning with Julia.''}
\emph{Journal of Open Source Software}.
\url{https://doi.org/10.21105/joss.00602}.

\leavevmode\vadjust pre{\hypertarget{ref-innes2019differentiable}{}}%
Innes, Mike, Alan Edelman, Keno Fischer, Chris Rackauckas, Elliot Saba,
Viral B Shah, and Will Tebbutt. 2019. {``A Differentiable Programming
System to Bridge Machine Learning and Scientific Computing.''}
\emph{arXiv Preprint arXiv:1907.07587}.

\leavevmode\vadjust pre{\hypertarget{ref-jones2012building}{}}%
Jones, CE, JA Edgar, JI Selvage, and H Crook. 2012. {``Building Complex
Synthetic Models to Evaluate Acquisition Geometries and Velocity
Inversion Technologies.''} In \emph{74th EAGE Conference and Exhibition
Incorporating EUROPEC 2012}, cp--293. European Association of
Geoscientists \& Engineers.

\leavevmode\vadjust pre{\hypertarget{ref-SegyIO}{}}%
Lensink, Keegan, Henryk Modzelewski, Mathias Louboutin, yzhang3198, and
Ziyi (Francis) Yin. 2023. \emph{Slimgroup/SegyIO.jl: V0.8.3} (version
v0.8.3). Zenodo. \url{https://doi.org/10.5281/zenodo.7502671}.

\leavevmode\vadjust pre{\hypertarget{ref-li2020coupled}{}}%
Li, Dongzhuo, Kailai Xu, Jerry M Harris, and Eric Darve. 2020.
{``Coupled Time-Lapse Full-Waveform Inversion for Subsurface Flow
Problems Using Intrusive Automatic Differentiation.''} \emph{Water
Resources Research} 56 (8): e2019WR027032.

\leavevmode\vadjust pre{\hypertarget{ref-li2020fourier}{}}%
Li, Zongyi, Nikola Kovachki, Kamyar Azizzadenesheli, Burigede Liu,
Kaushik Bhattacharya, Andrew Stuart, and Anima Anandkumar. 2020.
{``Fourier Neural Operator for Parametric Partial Differential
Equations.''} \url{https://arxiv.org/abs/2010.08895}.

\leavevmode\vadjust pre{\hypertarget{ref-lie2021advanced}{}}%
Lie, Knut-Andreas, and Olav Møyner. 2021. \emph{Advanced Modelling with
the MATLAB Reservoir Simulation Toolbox}. Cambridge University Press.

\leavevmode\vadjust pre{\hypertarget{ref-lin2015IIPFWIsdh}{}}%
Lin, Tim T. Y., and Felix J. Herrmann. 2015. {``The Student-Driven HPC
Environment at SLIM.''}
\url{https://slim.gatech.edu/Publications/Public/Conferences/IIPFWI/lin2015IIPFWIsdh/lin2015IIPFWIsdh_pres.pdf}.

\leavevmode\vadjust pre{\hypertarget{ref-louboutin2022enabling}{}}%
Louboutin, Mathias, and F Herrmann. 2022. {``Enabling Wave-Based
Inversion on GPUs with Randomized Trace Estimation.''} In \emph{83rd
EAGE Annual Conference \& Exhibition}, 2022:1--5. 1. EAGE Publications
BV.

\leavevmode\vadjust pre{\hypertarget{ref-louboutin2023rte}{}}%
Louboutin, Mathias, and Felix J. Herrmann. 2023. {``Wave-Based Inversion
at Scale on GPUs with Randomized Trace Estimation.''}
\url{https://slim.gatech.edu/Publications/Public/Submitted/2023/louboutin2023rte/paper.html}.

\leavevmode\vadjust pre{\hypertarget{ref-louboutin2022ais}{}}%
Louboutin, Mathias, Philipp Witte, Ali Siahkoohi, Gabrio Rizzuti, Ziyi
Yin, Rafael Orozco, and Felix J. Herrmann. 2022. {``Accelerating
Innovation with Software Abstractions for Scalable Computational
Geophysics.''} In \emph{Second International Meeting for Applied
Geoscience \&Amp; Energy}, 1482--86.
\url{https://doi.org/10.1190/image2022-3750561.1}.

\leavevmode\vadjust pre{\hypertarget{ref-judi}{}}%
Louboutin, Mathias, Philipp Witte, Ziyi Yin, Henryk Modzelewski, Kerim,
Carlos da Costa, and Peterson Nogueira. 2023. \emph{Slimgroup/JUDI.jl:
V3.2.3} (version v3.2.3). Zenodo.
\url{https://doi.org/10.5281/zenodo.7785440}.

\leavevmode\vadjust pre{\hypertarget{ref-JUDI4Cloud}{}}%
Louboutin, Mathias, Ziyi Yin, and Felix J. Herrmann. 2022a.
\emph{Slimgroup/JUDI4Cloud.jl: FIrst Public Release} (version v0.2.1).
Zenodo. \url{https://doi.org/10.5281/zenodo.6386831}.

\leavevmode\vadjust pre{\hypertarget{ref-SlimOptim}{}}%
---------. 2022b. \emph{Slimgroup/SlimOptim.jl: V0.2.0} (version
v0.2.0). Zenodo. \url{https://doi.org/10.5281/zenodo.7019463}.

\leavevmode\vadjust pre{\hypertarget{ref-devito-api}{}}%
Louboutin, M., M. Lange, F. Luporini, N. Kukreja, P. A. Witte, F. J.
Herrmann, P. Velesko, and G. J. Gorman. 2019. {``Devito (V3.1.0): An
Embedded Domain-Specific Language for Finite Differences and Geophysical
Exploration.''} \emph{Geoscientific Model Development} 12 (3): 1165--87.
\url{https://doi.org/10.5194/gmd-12-1165-2019}.

\leavevmode\vadjust pre{\hypertarget{ref-devito-compiler}{}}%
Luporini, Fabio, Mathias Louboutin, Michael Lange, Navjot Kukreja,
Philipp Witte, Jan Hückelheim, Charles Yount, Paul H. J. Kelly, Felix J.
Herrmann, and Gerard J. Gorman. 2020. {``Architecture and Performance of
Devito, a System for Automated Stencil Computation.''} \emph{ACM Trans.
Math. Softw.} 46 (1). \url{https://doi.org/10.1145/3374916}.

\leavevmode\vadjust pre{\hypertarget{ref-sympy}{}}%
Meurer, Aaron, Christopher P. Smith, Mateusz Paprocki, Ondřej Čertík,
Sergey B. Kirpichev, Matthew Rocklin, AMiT Kumar, et al. 2017. {``SymPy:
Symbolic Computing in Python.''} \emph{PeerJ Computer Science} 3
(January): e103. \url{https://doi.org/10.7717/peerj-cs.103}.

\leavevmode\vadjust pre{\hypertarget{ref-JOLI}{}}%
Modzelewski, Henryk, Mathias Louboutin, Ziyi Yin, Daniel Karrasch, and
Rafael Orozco. 2023. \emph{Slimgroup/JOLI.jl: V0.8.5} (version v0.8.5).
Zenodo. \url{https://doi.org/10.5281/zenodo.7752660}.

\leavevmode\vadjust pre{\hypertarget{ref-JutulDarcy}{}}%
Møyner, Olav, and Grant Bruer. 2023. \emph{Sintefmath/JutulDarcy.jl:
V0.2.2} (version v0.2.2). Zenodo.
\url{https://doi.org/10.5281/zenodo.7775738}.

\leavevmode\vadjust pre{\hypertarget{ref-Jutul}{}}%
Møyner, Olav, Martin Johnsrud, Halvor Møll Nilsen, Xavier Raynaud,
Kjetil Olsen Lye, and Ziyi Yin. 2023. \emph{Sintefmath/Jutul.jl: V0.2.5}
(version v0.2.5). Zenodo. \url{https://doi.org/10.5281/zenodo.7775759}.

\leavevmode\vadjust pre{\hypertarget{ref-orozco2023MIDLanf}{}}%
Orozco, Rafael, Mathias Louboutin, Ali Siahkoohi, Gabrio Rizzuti,
Tristan van Leeuwen, and Felix J. Herrmann. 2023. {``Amortized
Normalizing Flows for Transcranial Ultrasound with Uncertainty
Quantification,''} March.
\url{https://openreview.net/forum?id=LoJG-lUIlk}.

\leavevmode\vadjust pre{\hypertarget{ref-orozco2021photoacoustic}{}}%
Orozco, Rafael, Ali Siahkoohi, Gabrio Rizzuti, Tristan van Leeuwen, and
Felix J. Herrmann. 2021. {``Photoacoustic Imaging with Conditional
Priors from Normalizing Flows.''}
\url{https://openreview.net/forum?id=woi1OTvROO1}.

\leavevmode\vadjust pre{\hypertarget{ref-adjoint}{}}%
---------. 2023. {``{Adjoint operators enable fast and amortized machine
learning based Bayesian uncertainty quantification}.''} In \emph{Medical
Imaging 2023: Image Processing}, edited by Olivier Colliot and Ivana
Išgum, 12464:124641L. International Society for Optics; Photonics; SPIE.
\url{https://doi.org/10.1117/12.2651691}.

\leavevmode\vadjust pre{\hypertarget{ref-ricevector}{}}%
Padula, Anthony D., Shannon D. Scott, and William W. Symes. 2009. {``A
Software Framework for Abstract Expression of Coordinate-Free Linear
Algebra and Optimization Algorithms.''} \emph{ACM Trans. Math. Softw.}
36 (2). \url{https://doi.org/ricevector}.

\leavevmode\vadjust pre{\hypertarget{ref-peters2019algorithms}{}}%
Peters, Bas, and Felix J Herrmann. 2019. {``Algorithms and Software for
Projections onto Intersections of Convex and Non-Convex Sets with
Applications to Inverse Problems.''} \emph{arXiv Preprint
arXiv:1902.09699}.

\leavevmode\vadjust pre{\hypertarget{ref-SetIntersectionProjection}{}}%
Peters, Bas, Mathias Louboutin, and Henryk Modzelewski. 2022.
\emph{Slimgroup/SetIntersectionProjection.jl: V0.2.4} (version v0.2.4).
Zenodo. \url{https://doi.org/10.5281/zenodo.7257913}.

\leavevmode\vadjust pre{\hypertarget{ref-rasmussen2021open}{}}%
Rasmussen, Atgeirr Flø, Tor Harald Sandve, Kai Bao, Andreas Lauser,
Joakim Hove, Bård Skaflestad, Robert Klöfkorn, et al. 2021. {``The Open
Porous Media Flow Reservoir Simulator.''} \emph{Computers \& Mathematics
with Applications} 81: 159--85.

\leavevmode\vadjust pre{\hypertarget{ref-firedrake}{}}%
Rathgeber, Florian, David A. Ham, Lawrence Mitchell, Michael Lange,
Fabio Luporini, Andrew T. T. Mcrae, Gheorghe-Teodor Bercea, Graham R.
Markall, and Paul H. J. Kelly. 2016. {``Firedrake: Automating the Finite
Element Method by Composing Abstractions.''} \emph{ACM Trans. Math.
Softw.} 43 (3). \url{https://doi.org/10.1145/2998441}.

\leavevmode\vadjust pre{\hypertarget{ref-RAVASI2020100361}{}}%
Ravasi, Matteo, and Ivan Vasconcelos. 2020. {``PyLops---a
Linear-Operator Python Library for Scalable Algebra and Optimization.''}
\emph{SoftwareX} 11: 100361.
https://doi.org/\url{https://doi.org/10.1016/j.softx.2019.100361}.

\leavevmode\vadjust pre{\hypertarget{ref-rezende2015variational}{}}%
Rezende, Danilo, and Shakir Mohamed. 2015. {``Variational Inference with
Normalizing Flows.''} In \emph{Proceedings of Machine Learning
Research}, 37:1530--38. Proceedings of Machine Learning Research. PMLR.
\url{http://proceedings.mlr.press/v37/rezende15.html}.

\leavevmode\vadjust pre{\hypertarget{ref-ringrose2020store}{}}%
Ringrose, Philip. 2020. \emph{How to Store CO2 Underground: Insights
from Early-Mover CCS Projects}. Springer.

\leavevmode\vadjust pre{\hypertarget{ref-rizzuti2020dfw}{}}%
Rizzuti, Gabrio, Mathias Louboutin, Rongrong Wang, and Felix J.
Herrmann. 2021. {``A Dual Formulation of Wavefield Reconstruction
Inversion for Large-Scale Seismic Inversion.''} \emph{Geophysics} 86
(6): 1ND--Z3. \url{https://doi.org/10.1190/geo2020-0743.1}.

\leavevmode\vadjust pre{\hypertarget{ref-rizzuti2020SEGpub}{}}%
Rizzuti, Gabrio, Ali Siahkoohi, Philipp A. Witte, and Felix J. Herrmann.
2020. {``Parameterizing Uncertainty by Deep Invertible Networks, an
Application to Reservoir Characterization.''} In \emph{90th Annual
International Meeting}, 1541--45. Society of Exploration Geophysicists;
Expanded Abstracts. \url{https://doi.org/10.1190/segam2020-3428150.1}.

\leavevmode\vadjust pre{\hypertarget{ref-python}{}}%
Rossum, Guido van, and Fred L. Drake. 2009. \emph{Python 3 Reference
Manual}. Scotts Valley, CA: CreateSpace.

\leavevmode\vadjust pre{\hypertarget{ref-Settgast_GEOSX_2022}{}}%
Settgast, Randolph Richard, Benjamin Curtice Corbett, Sergey Klevtsov,
Francois Hamon, Christopher Sherman, Matteo Cusini, Thomas Gazzola, et
al. 2022. {``{GEOSX}.''} \url{https://doi.org/10.5281/zenodo.7151032}.

\leavevmode\vadjust pre{\hypertarget{ref-siahkoohi2021Seglbe}{}}%
Siahkoohi, Ali, and Felix J. Herrmann. 2021. {``Learning by Example:
Fast Reliability-Aware Seismic Imaging with Normalizing Flows.''} In
\emph{First International Meeting for Applied Geoscience {\&} Energy},
1580--85. Society of Exploration Geophysicists; Expanded Abstracts.
\url{https://doi.org/10.1190/segam2021-3581836.1}.

\leavevmode\vadjust pre{\hypertarget{ref-siahkoohi2020EAGEdlb}{}}%
Siahkoohi, Ali, Gabrio Rizzuti, and Felix J. Herrmann. 2020a. {``A
Deep-Learning Based Bayesian Approach to Seismic Imaging and Uncertainty
Quantification.''} In \emph{82nd EAGE Conference and Exhibition}.
Extended Abstracts. \url{https://doi.org/10.3997/2214-4609.202010770}.

\leavevmode\vadjust pre{\hypertarget{ref-siahkoohi2020SEGuqi}{}}%
---------. 2020b. {``{Uncertainty quantification in imaging and
automatic horizon tracking---a Bayesian deep-prior based approach}.''}
In \emph{90th Annual International Meeting}, 1636--40. Society of
Exploration Geophysicists; Expanded Abstracts.
\url{https://doi.org/10.1190/segam2020-3417560.1}.

\leavevmode\vadjust pre{\hypertarget{ref-siahkoohi2020SEGwdp}{}}%
---------. 2020c. {``Weak Deep Priors for Seismic Imaging.''} In
\emph{90th Annual International Meeting}, 2998--3002. Society of
Exploration Geophysicists; Expanded Abstracts.
\url{https://doi.org/10.1190/segam2020-3417568.1}.

\leavevmode\vadjust pre{\hypertarget{ref-siahkoohiGEOdbif}{}}%
---------. 2022. {``{D}eep {B}ayesian Inference for Seismic Imaging with
Tasks.''} \emph{Geophysics} 87 (5): S281--302.
\url{https://doi.org/10.1190/geo2021-0666.1}.

\leavevmode\vadjust pre{\hypertarget{ref-siahkoohi2020ABIpto}{}}%
Siahkoohi, Ali, Gabrio Rizzuti, Mathias Louboutin, Philipp Witte, and
Felix J. Herrmann. 2021. {``Preconditioned Training of Normalizing Flows
for Variational Inference in Inverse Problems.''}
\url{https://openreview.net/pdf?id=P9m1sMaNQ8T}.

\leavevmode\vadjust pre{\hypertarget{ref-doi:10.1190ux2fgeo2022-0472.1}{}}%
Siahkoohi, Ali, Gabrio Rizzuti, Rafael Orozco, and Felix J. Herrmann.
2023. {``Reliable Amortized Variational Inference with Physics-Based
Latent Distribution Correction.''} \emph{Geophysics} 88 (3).
\url{https://doi.org/10.1190/geo2022-0472.1}.

\leavevmode\vadjust pre{\hypertarget{ref-dasilva2017uls}{}}%
Silva, Curt Da, and Felix J. Herrmann. 2019. {``A Unified 2D/3D Large
Scale Software Environment for Nonlinear Inverse Problems.''} \emph{ACM
Transactions on Mathematical Software}.
\url{https://slim.gatech.edu/Publications/Public/Journals/ACMTOMS/2019/dasilva2017uls/dasilva2017uls.html}.

\leavevmode\vadjust pre{\hypertarget{ref-song2020score}{}}%
Song, Yang, Jascha Sohl-Dickstein, Diederik P Kingma, Abhishek Kumar,
Stefano Ermon, and Ben Poole. 2020. {``Score-Based Generative Modeling
Through Stochastic Differential Equations.''} \emph{arXiv Preprint
arXiv:2011.13456}.

\leavevmode\vadjust pre{\hypertarget{ref-sun2010iwave}{}}%
Sun, Dong, and William W Symes. 2010. {``IWAVE Implementation of Adjoint
State Method.''} Tech. Rep. 10-06, Department of Computational; Applied
Mathematics, Rice University, Houston, Texas, USA.
\url{https://pdfs.semanticscholar.org/6c17/cfe41b76f6b745c435891ea6ba6f4e2c2dbf.pdf}.

\leavevmode\vadjust pre{\hypertarget{ref-cofii}{}}%
Washbourne, John, Sam Kaplan, Miguel Merino, Uwe Albertin, Anusha Sekar,
Chris Manuel, Scott Mishra, Matthew Chenette, and Alex Loddoch. 2021.
{``Chevron Optimization Framework for Imaging and Inversion (COFII) --
an Open Source and Cloud Friendly Julia Language Framework for Seismic
Modeling and Inversion.''} In \emph{First International Meeting for
Applied Geoscience \&Amp; Energy Expanded Abstracts}, 792--96.
\url{https://doi.org/10.1190/segam2021-3594362.1}.

\leavevmode\vadjust pre{\hypertarget{ref-wen2022u}{}}%
Wen, Gege, Zongyi Li, Kamyar Azizzadenesheli, Anima Anandkumar, and
Sally M Benson. 2022. {``{U-FNO}---an Enhanced Fourier Neural
Operator-Based Deep-Learning Model for Multiphase Flow.''}
\emph{Advances in Water Resources} 163: 104180.

\leavevmode\vadjust pre{\hypertarget{ref-wen2023real}{}}%
Wen, Gege, Zongyi Li, Qirui Long, Kamyar Azizzadenesheli, Anima
Anandkumar, and Sally Benson. 2023. {``Real-Time High-Resolution CO2
Geological Storage Prediction Using Nested Fourier Neural Operators.''}
\emph{Energy \& Environmental Science}.

\leavevmode\vadjust pre{\hypertarget{ref-frames_catherine_white_2023_7628788}{}}%
White, Frames Catherine, Michael Abbott, Miha Zgubic, Jarrett Revels,
Seth Axen, Alex Arslan, Simeon Schaub, et al. 2023.
{``JuliaDiff/ChainRules.jl: V1.47.0.''} Zenodo.
\url{https://doi.org/10.5281/zenodo.7628788}.

\leavevmode\vadjust pre{\hypertarget{ref-frames_catherine_white_2022_7107911}{}}%
White, Frames Catherine, Miha Zgubic, Michael Abbott, Jarrett Revels,
Nick Robinson, Alex Arslan, David Widmann, et al. 2022.
{``JuliaDiff/ChainRulesCore.jl: V1.15.6.''} Zenodo.
\url{https://doi.org/10.5281/zenodo.7107911}.

\leavevmode\vadjust pre{\hypertarget{ref-witte2022sciai4industry}{}}%
Witte, Philipp A., Russell J. Hewett, Kumar Saurabh, AmirHossein
Sojoodi, and Ranveer Chandra. 2022. {``SciAI4Industry -- Solving PDEs
for Industry-Scale Problems with Deep Learning.''}
\url{https://arxiv.org/abs/2211.12709}.

\leavevmode\vadjust pre{\hypertarget{ref-witte2018alf}{}}%
Witte, Philipp A., Mathias Louboutin, Navjot Kukreja, Fabio Luporini,
Michael Lange, Gerard J. Gorman, and Felix J. Herrmann. 2019. {``A
Large-Scale Framework for Symbolic Implementations of Seismic Inversion
Algorithms in Julia.''} \emph{Geophysics} 84 (3): F57--71.
\url{https://doi.org/10.1190/geo2018-0174.1}.

\leavevmode\vadjust pre{\hypertarget{ref-witte2018cls}{}}%
Witte, Philipp A., Mathias Louboutin, Fabio Luporini, Gerard J. Gorman,
and Felix J. Herrmann. 2019. {``Compressive Least-Squares Migration with
on-the-Fly Fourier Transforms.''} \emph{Geophysics} 84 (5): R655--72.
\url{https://doi.org/10.1190/geo2018-0490.1}.

\leavevmode\vadjust pre{\hypertarget{ref-InvertibleNetworks}{}}%
Witte, Philipp, Mathias Louboutin, Rafael Orozco, Gabrio Rizzuti, Ali
Siahkoohi, Felix Herrmann, Bas Peters, Páll Haraldsson, and Ziyi Yin.
2023. \emph{Slimgroup/InvertibleNetworks.jl: V2.2.4} (version v2.2.4).
Zenodo. \url{https://doi.org/10.5281/zenodo.7693048}.

\leavevmode\vadjust pre{\hypertarget{ref-yang2020tdsp}{}}%
Yang, Mengmeng, Zhilong Fang, Philipp A. Witte, and Felix J. Herrmann.
2020. {``Time-Domain Sparsity Promoting Least-Squares Reverse Time
Migration with Source Estimation.''} \emph{Geophysical Prospecting} 68
(9): 2697--2711. \url{https://doi.org/10.1111/1365-2478.13021}.

\leavevmode\vadjust pre{\hypertarget{ref-yin2022TLEdgc}{}}%
Yin, Ziyi, Huseyin Tuna Erdinc, Abhinav Prakash Gahlot, Mathias
Louboutin, and Felix J. Herrmann. 2023. {``Derisking Geologic Carbon
Storage from High-Resolution Time-Lapse Seismic to Explainable Leakage
Detection.''} \emph{The Leading Edge} 42 (1): 69--76.
\url{https://doi.org/10.1190/tle42010069.1}.

\leavevmode\vadjust pre{\hypertarget{ref-JutulDarcyRules}{}}%
Yin, Ziyi, and Mathias Louboutin. 2023.
\emph{Slimgroup/JutulDarcyRules.jl: V0.2.4} (version v0.2.4). Zenodo.
\url{https://doi.org/10.5281/zenodo.7762154}.

\leavevmode\vadjust pre{\hypertarget{ref-yin2021SEGcts}{}}%
Yin, Ziyi, Mathias Louboutin, and Felix J. Herrmann. 2021.
{``Compressive Time-Lapse Seismic Monitoring of Carbon Storage and
Sequestration with the Joint Recovery Model.''}
\url{https://doi.org/10.1190/segam2021-3569087.1}.

\leavevmode\vadjust pre{\hypertarget{ref-yin2020SEGesi}{}}%
Yin, Ziyi, Rafael Orozco, Philipp A. Witte, Mathias Louboutin, Gabrio
Rizzuti, and Felix J. Herrmann. 2020. {``Extended Source Imaging --- a
Unifying Framework for Seismic and Medical Imaging.''}
\url{https://doi.org/10.1190/segam2020-3426999.1}.

\leavevmode\vadjust pre{\hypertarget{ref-yin2022learned}{}}%
Yin, Ziyi, Ali Siahkoohi, Mathias Louboutin, and Felix J. Herrmann.
2022. {``Learned Coupled Inversion for Carbon Sequestration Monitoring
and Forecasting with Fourier Neural Operators.''}
\url{https://doi.org/10.1190/image2022-3722848.1}.

\leavevmode\vadjust pre{\hypertarget{ref-zhang2020seismic}{}}%
Zhang, Xin, and Andrew Curtis. 2020. {``Seismic Tomography Using
Variational Inference Methods.''} \emph{Journal of Geophysical Research:
Solid Earth} 125 (4): e2019JB018589.

\leavevmode\vadjust pre{\hypertarget{ref-zhang2021bayesian2}{}}%
---------. 2021. {``Bayesian Geophysical Inversion Using Invertible
Neural Networks.''} \emph{Journal of Geophysical Research: Solid Earth}
126 (7): e2021JB022320.

\leavevmode\vadjust pre{\hypertarget{ref-zhao2020bayesian}{}}%
Zhao, Xuebin, Andrew Curtis, and Xin Zhang. 2021. {``{Bayesian seismic
tomography using normalizing flows}.''} \emph{Geophysical Journal
International} 228 (1): 213--39.
\url{https://doi.org/10.1093/gji/ggab298}.

\end{CSLReferences}

\end{document}